\documentclass{aa}  

\usepackage{graphicx}
\usepackage{txfonts}
\usepackage{lipsum}
\usepackage[caption=false]{subfig}
\usepackage{lscape}             
\usepackage{placeins}
\usepackage{natbib}
\usepackage{comment}
\usepackage[thinc]{esdiff}
\usepackage[]{hyperref}

\begin{document}

   \title{Connecting coronal loop scaling laws and kink wave heating}

   \subtitle{An analytical and numerical study}

   \author{L. Banović\inst{1}\fnmsep\corrauth{luka.banovic@kuleuven.be}
        \and T. Van Doorsselaere\inst{1}\email{tom.vandoorsselaere@kuleuven.be}
        \and M. McMurdo\inst{1}\email{max.mcmurdo@kuleuven.be}
        \and H. Washinoue\inst{2}\email{haruka.washinoue@riken.jp}
        \and A. Krupka\inst{1}\email{anna.krupka@kuleuven.be}
        }

   \institute{Centre for mathematical Plasma Astrophysics, Department of Mathematics, KU Leuven, Celestijnenlaan 200B Bus 2400, 3001 Leuven, Belgium
   \and Pioneering Research Institute, RIKEN, 2-1, Hirosawa, Wako, Saitama 351-0198, Japan\\}

   \date{Received date / Accepted date}

  \abstract
   {Turbulent dissipation of kink waves has been proposed as one of the mechanisms responsible for the observed high coronal temperatures. This hypothesis can be tested by examining the effect of this heating mechanism on coronal loops.}
   {We aim to investigate coronal loop static and dynamic behaviour under the assumption that the heating in the loop is primarily provided by kink wave dissipation. We want to provide estimates of the kink wave energy required to sustain a coronal loop against radiative cooling, as well as scaling laws relating kink wave energy density to other loop parameters such as pressure and loop length.}
   {The heating term is borrowed from the recently developed \mbox{UAWSoM} model. Loop modelling is done in 1D using three different methods. First, we do an analytical study following the principle first set in the RTV scaling laws. Two numerical approaches are also examined: one which is based on the setup that was used to derive RTVS scaling laws, and the other which uses \mbox{UAWSoM}.}
   {We provide estimates of density, temperature, pressure, kink wave heating rate and kink wave energy density, including scaling relationships between different loop quantities. Estimates of kink wave energy density needed to heat the loop to 1 MK correspond to maximum velocity amplitudes of around 35 km s$^{-1}$ near the loop apex. Loop radius, density contrast and filling factor are found to affect the predictions significantly. The results of the two numerical methods agree to a high degree. The analytical method predicts that $W_\textrm{k} \propto PL^{-2/3}$, while the two numerical methods show a wider range of scaling exponents which depend on the location in the loop. UAWSoM simulations suggest that loops may exhibit dynamic behaviour for a wide range of loop conditions, which may include thermal non-equilibrium (TNE) cycles and periods of emerging and dissipating slow waves.}
   {A comprehensive picture of how coronal loops may be affected by uniturbulent heating is provided. Estimates of kink wave energy density and scaling laws provide observational constraints on the importance of the proposed kink wave heating mechanism. Current preliminary results suggest that the uniturbulent kink wave heating may be a significant source of energy and heat in coronal loops, though more investigation is needed. This study also serves as a stepping stone in the further development of the \mbox{UAWSoM} model.}


   \maketitle
   \nolinenumbers

\section{Introduction}

Coronal loop scaling laws, establishing the relationship between pressure, loop length, maximum temperature and heating rate, have been crucial in studying characteristics of loop plasma, aiming to explain the observations and possibly help in constraining heating mechanisms responsible for observed high plasma temperatures of the order of several MK. First introduced by \citet{Rosner1978} (and today known as RTV scaling laws), the authors analytically derived the two scaling laws for loops under the assumptions of constant pressure and uniform heating. Many extensions of the methodology introduced in this seminal work have since been put forward. \citet{Hood1979} analysed the case of uniform heating per unit mass and assessed the effects of changing radial magnetic field and magnetic twist. \citet{Vesecky1979} included gravitational stratification and expansion of the loop with height, which required them to solve the underlying set of equations behind the scaling laws numerically. \citet{Serio1981} added to this by replacing uniform with an exponentially decaying heating rate. \citet{Aschwanden2002} extended this work further by providing more accurate analytical approximations to the numerical solutions. \citet{Martens2010} analytically derived the temperature profiles and two scaling laws in the case of constant pressure and non-uniform heating. This non-uniform heating is assumed to follow a scaling relationship with respect to temperature, which is the key assumption that enables the derivation of the scaling laws. More recently, different thermal conduction mechanisms \citep{Bradshaw2019, Asgari-Targhi2025}, dynamic coronal loops \citep{Bradshaw2020}, loops in environments with different metallicities \citep{Washinoue2023}, and various combinations thereof have been considered \citep{Emslie2022, Dai2024}. The scaling laws have been matched with observations with varying success \citep[e.g.][]{Rosner1978, Kano1995, Aschwanden2000, MacCormack2020, Bennedik2026submitted}, however they remain useful as an important baseline in coronal loop studies \citep{Reale2014}.

One of the key ingredients in every coronal loop model is the assumed underlying heating mechanism, which counteracts the continuous loss of energy and energy redistribution through radiative cooling and thermal conduction. The debate about the prevalent heating mechanism(s) is still ongoing, with recent overviews by \citet{Cranmer2019} and \citet{VanDoorsselaere2020}. In particular, several heating mechanisms based on the wave-plasma interactions have been investigated, such as phase mixing \citep[][and many more]{Heyvaerts1983, Cargill2016, Pagano2017, Howson2020, McMurdo2025}, energy dissipation through transverse wave induced Kelvin-Helmholtz rolls \citep[e.g.][]{Terradas2008, Karampelas2017, Karampelas2024, Shi2021, Zhong2026}, and turbulent dissipation of Alfvén waves \citep[e.g.][]{Hollweg1986, Verdini2007, vanBallegooijen2017, Shi2024}. In the context of RTV scaling laws and their derivatives, uniform heating and heating profiles which exponentially decrease with height are most commonly assumed for their simplicity. 

This work focuses on one of the proposed mechanisms, namely the dissipation of kink waves through uniturbulence. Investigations by \citet{Magyar2017, Magyar2019} have shown that kink waves propagating only in a single direction can induce turbulence in the presence of density inhomogeneities. While the energy dissipation of the Alfvén waves has already been examined and included in global coronal models such as \mbox{AWSoM} or \mbox{WTD} \citep{vanderHolst2014, Downs2016, Shoda2019, Reville2020}, this has not been the case for the uniturbulent dissipation until recently. This recent advancement was made possible thanks to the novel wave-aligned formulation of magnetohydrodynamics (MHD) equations in terms of the Q-variables:
\begin{equation}
 \mathbf{Q}^{\pm}= \vec{\varv}\pm\alpha \vec{B},
\end{equation}
where the parameter $\alpha$ is determined by the local plasma state and the selected wave branch \citep{VanDoorsselaere2024}. While the classical Elsässer variables \citep{Elsasser1950} are intrinsically associated with Alfvénic propagation (and can be recovered with the appropriate choice of $\alpha$), the Q-variable formalism provides a wave-aligned representation applicable to a broader class of MHD disturbances, including magnetoacoustic and kink waves. The consistency of the Q-variable framework has also recently been demonstrated in the presence of MHD discontinuities by \citet{Krupka2026}, who derived the complete Rankine--Hugoniot conditions in Q-variables and demonstrated their exact equivalence to the classical ideal-MHD jump conditions. The Q-variable framework also provided the description of energy transport and nonlinear cascade of propagating kink waves, which proved to be a key step in the formalisation of the \mbox{UAWSoM} model \citep[Uniturbulence and Alfvén Wave Solar Model;][]{VanDoorsselaere2025}. Thus, in \mbox{UAWSoM}, the energy input from the dissipation of both the Alfvén and kink waves is taken into account, with the recent study by \citet{McMurdo2026} suggesting that energy provided by dissipating kink waves is capable of heating the lower corona and plays a more dominant role compared to Alfvén wave heating.

However, the viability of kink wave dissipation as a significant energy source in the corona is strongly dependent on how commonly they occur and how efficiently they transfer energy from the photosphere to the corona. In the past 25 years, many observations of the transverse oscillations have been recorded, starting with \citet{Aschwanden1999} and \citet{Nakariakov1999}, confirming their omnipresent nature in the corona. Current energy flux estimates suggest that enough energy may be transported by transverse oscillations into the corona to sustain it against radiative cooling \citep{Lim2024, Shrivastav2024, Morton2026}. However, observational constraints are still an important limiting factor in providing a definitive conclusion as to how much energy is carried by transverse oscillations into the corona. This is a large factor of uncertainty in the \mbox{UAWSoM} model (and both \mbox{UAWSoM} and AWSoM models with respect to the Alfvén waves), where the model predictions of plasma parameters are closely tied to the assumed energy flux carried by coronal waves. These uncertainties become even more pronounced when such global coronal models are used to model environments of stars other than the Sun \citep[e.g.][]{Vidotto2023}. 

In this work, we aim to provide better constraints on the main ingredient of the kink wave heating term put forward in \citet{VanDoorsselaere2025}, which is the kink wave energy density. We do this by investigating the coronal loop behaviour under the assumption of it being heated primarily by uniturbulent kink wave dissipation. Three methods are considered: the analytical setting of RTV scaling laws in the context of non-uniform heating by \citet{Martens2010}, numerical solutions of hydrostatic equations based on the work of \citet{Serio1981} and \mbox{UAWSoM} simulations of coronal loops in one dimension (1D). This allows for a thorough analysis of similarities and discrepancies between the different methods characterised by varying assumptions and computational complexity. Apart from only providing predictions of the energy content of the kink waves needed and the resulting coronal loop behaviour, we also focus on the results of the \mbox{UAWSoM} model in a closed-field setting, which has only been investigated very recently \citep{McMurdo2026b}. A thorough analysis of results in this setting is important if \mbox{UAWSoM} is to be reliably used in global coronal modelling.

The paper is organised as follows. In Section~\ref{sect:methods}, we introduce the framework in which the investigation is done, together with the analytical and numerical methods employed. Results and predictions of these methods are presented and discussed in Section~\ref{sect:results}, with the emphasis on the differences between the applied methods. Finally, the conclusions are summarised in Section~\ref{sect:conclusion} with proposals of possible future research avenues.

\section{Methods}\label{sect:methods}

\subsection{Context, background and notation}

\begin{figure}[ht]
    \centering
    \includegraphics[width=3.5cm]{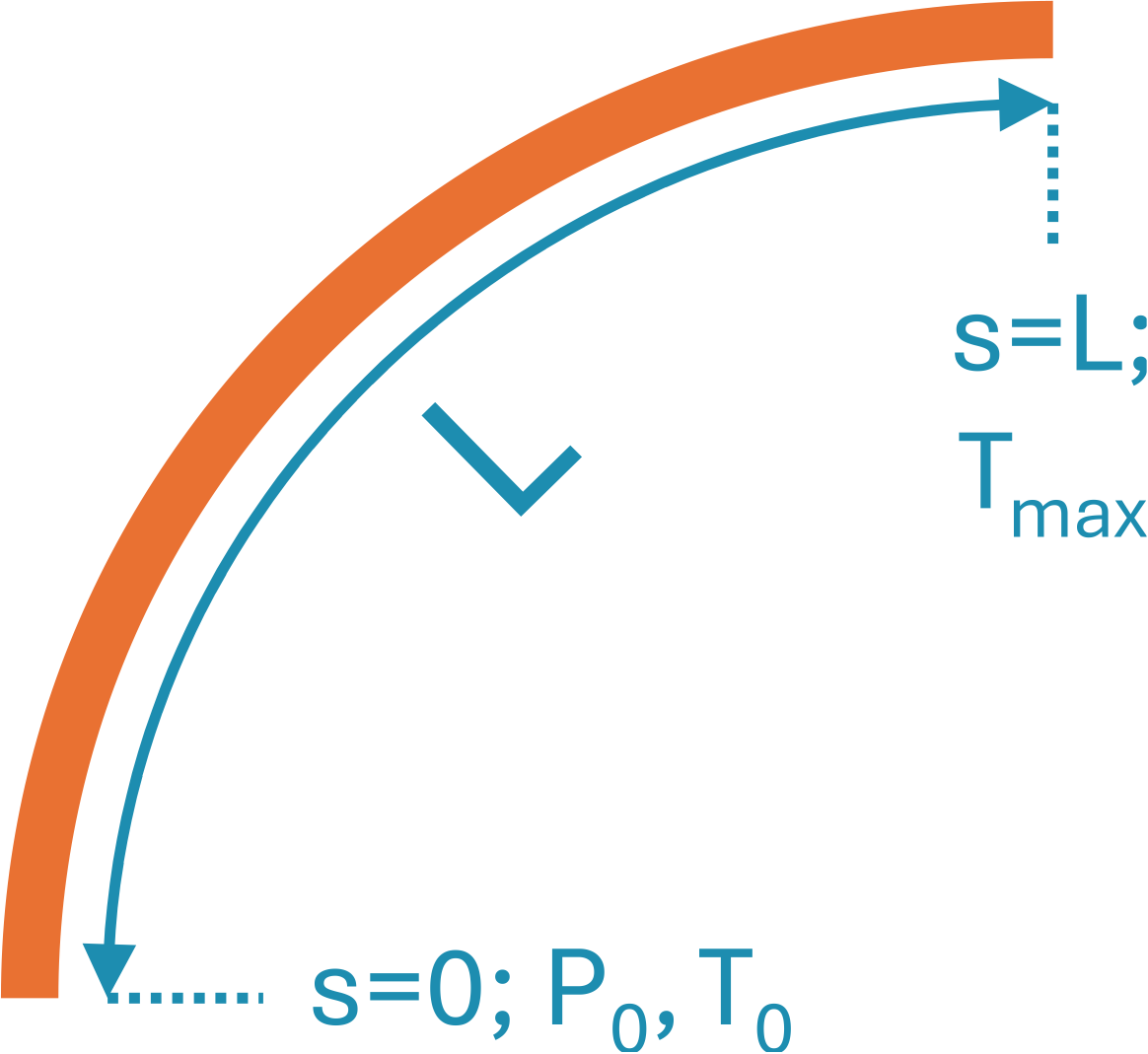}
    \caption{Loop schematics with some of the key parameters denoted. Bottom of the loop corresponds to $s=0$; top to $s=L$.}
    \label{fig:loop_schematics}
\end{figure}

Throughout this study, we model a coronal loop as a 1D structure along the loop. The large-scale geometry of one half of the semicircular loop is schematically depicted in Figure~\ref{fig:loop_schematics}. Several variables are introduced in this sketch: $s$ represents the spatial variable along the loop, $L$ the loop half-length, $P_0$ and $T_0$ are pressure and temperature at the bottom and $T_\textrm{max}$ is the maximum temperature, which for the steady-case models is assumed to be at the loop apex. In this 1D setting, we also assume that there is no energy transport through the sides of the loops.

Even though we model a coronal loop only in 1D, we take into account the small-scale structure across the loop in the manner consistent with the one shown in Fig.~1 in \citet{VanDoorsselaere2014}. Specifically, we assume that the coronal loop consists of an array of dense threads embedded in a lower-density exterior plasma. We denote the interior and exterior density by $\rho_\textrm{i}$ and $\rho_\textrm{e}$, respectively, and define the density contrast as $\zeta = \rho_\textrm{i}/\rho_\textrm{e}$. The filling factor $f$ denotes the ratio of the volume filled by the dense plumes compared to the volume of the whole loop. This means that the averaged cross-section density of the whole loop is equal to $\rho=f\rho_\textrm{i} + (1-f)\rho_\textrm{e}$ \citep[Eq.~(8) in][]{McMurdo2026}. We also denote the typical radius of a coronal loop by $R$. While we consider density variations inside the loop, the temperature is assumed not to differ between the dense plumes and the less dense plasma surrounding them.

The central equation in this work is arguably the energy equilibrium equation. In the simplest case, as proposed in \citet{Rosner1978}, the heating rate $E_\textrm{H}$ (in units of energy per unit volume and time), radiative losses $E_\textrm{R}$ and thermal conduction expressed in terms of the thermal conductive flux $F_\textrm{c}$ are in balance, meaning
\begin{equation}\label{eq:energy_equilibrium}
    E_\textrm{H} - E_\textrm{R} - \nabla\cdot F_\textrm{c} = 0.
\end{equation}
Optically thin radiative losses and Spitzer conductivity are considered, taking the form of
\begin{align}
    E_\textrm{R} &= \left(fn_\textrm{i}^2 + (1-f)n_\textrm{e}^2\right)\Lambda(T),\\
    F_\textrm{c} &= -\kappa T^{5/2} T', \label{eq:Fc}
\end{align}
where $n_\textrm{i}$ and $n_\textrm{e}$ are respectively the number density inside and outside of the dense loop threads, $\Lambda(T)$ is the radiative loss function which depends on temperature, and $\kappa$ is the Spitzer conductivity constant. Here we also introduce the symbol $'$ to represent specifically the derivative of a function along the loop: $' = \mathrm{d}/\mathrm{d}s$. Under the assumption that the corona consists of fully ionised plasma consisting of hydrogen and helium ions and electrons, 
\begin{equation}
    n_{i,e}=\frac{\rho_{i, e}}{\mu m_\textrm{p}}, \label{eq:n_definition}
\end{equation}
where 
\begin{equation}
    \mu = 1+4A_{\textrm{He}}
\end{equation}
is expressed in terms of the abundance ratio $A_{\textrm{He}}$ between helium and hydrogen ions \citep{Hermans2021}. The expression for the radiative losses takes into account the small-scale density structure of the loop, meaning that the radiative cooling contributions from dense thin strands and the surrounding plasma of lower density are treated separately and then added together (assuming that both regions have equal temperature). This philosophy follows the treatment set in \citet{McMurdo2026}. The calculations there can also be used to express $E_\textrm{R}$ in terms of the average density $\rho$:
\begin{equation}
    E_\textrm{R} = \frac{1-f + f\zeta^2}{(1 - f + f\zeta)^{2}} \left(\frac{\rho}{\mu m_\textrm{p}}\right)^2 \Lambda(T). \label{eq:ER}
\end{equation}

In this work, we assume that the loop is primarily heated by the dissipation of kink waves. We use the analytical heating rate for propagating kink waves by \citet{VanDoorsselaere2025}:
\begin{align}
    E_\textrm{H}^\pm &= \frac{1}{L_{\perp, \textrm{VD}}\sqrt{\rho_\textrm{e}}}(W_\textrm{k}^\pm)^{3/2} \\
    &= \frac{1}{L_{\perp, \textrm{VD}}}\sqrt{\frac{1-f+f\zeta}{\rho}}(W_\textrm{k}^\pm)^{3/2} \\
    &= \frac{1}{L_{\perp, \textrm{VD}}} \sqrt{\frac{\nu k_\textrm{B}(1-f+f\zeta)}{\mu m_\textrm{p}} \frac{T}{P}} (W_\textrm{k}^\pm)^{3/2} \label{eq:EH_init}.
\end{align}
In the last equation, we used the appropriate expression for the ideal gas law, which accounts for a plasma containing both hydrogen and helium ions:
\begin{equation}
    P = \nu nk_\textrm{B}T, \label{eq:ideal_gas_law}
\end{equation}
where 
\begin{equation}
    \nu = 2+3A_{\textrm{He}}
\end{equation}
and $k_\textrm{B}$ is the Boltzmann constant. The $\pm$ sign represents kink waves propagating in two opposite directions along the coronal loop. The total heating rate $E_\textrm{H}$ is the sum of these two contributions. This heating rate is expressed in terms of the kink wave energy densities $W_\textrm{k}^\pm$, which have the units of energy per unit volume. We also introduce the total kink wave energy $\bar{W}_\textrm{k}$ using the relationship 
\begin{equation}\label{eq:Wk_tot_definition}
    \bar{W}_\textrm{k}^{3/2} = (W_\textrm{k}^+)^{3/2} + (W_\textrm{k}^-)^{3/2}.
\end{equation}
As a result, the total heating rate $E_\textrm{H}$ can be simply written as
\begin{equation}
    E_\textrm{H} = \frac{1}{L_{\perp, \textrm{VD}}} \sqrt{\frac{\nu k_\textrm{B}(1-f+f\zeta)}{\mu m_\textrm{p}} \frac{T}{P}} \bar{W}_\textrm{k}^{3/2}. \label{eq:EH_total}
\end{equation}
This expression will be used in the analytical part (Sect.~\ref{sect:methods_analytics}). $L_{\perp, \textrm{VD}}$ is the length scale of kink wave dissipation, which is given by
\begin{equation}\label{eq:L_perp}
    L_{\perp, \textrm{VD}} = \left(\frac{1}{R\sqrt{f\pi}}\sqrt{\frac{2}{5}}\frac{\zeta-1}{2}\frac{1-f^{5/2}}{(\zeta+1-f)^{3/2}}\right) ^{-1}.
\end{equation}

\subsection{Analytical approach}\label{sect:methods_analytics}

In the first step, we derive the scaling law for the kink wave energy density analytically. This is done using the framework set by \citet{Martens2010}, where the coronal loop scaling laws and temperature profile inside the loop are derived under the assumption of a static symmetric loop which is heated non-uniformly.

Apart from the loop being static and symmetric, several other assumptions have to be made. First, pressure is assumed to be constant along the loop, with the implication that the height of the loop top is much smaller than the pressure scale height in the corona. Next, we approximate the radiative loss function by a scaling law of the form 
\begin{equation}\label{eq:Lambda}
    \Lambda(T) = \chi T^\alpha,
\end{equation} with constants $\chi$ and $\alpha$ obtained by fitting the actual cooling curves with this scaling law, as in \citet{Martens2010} and \citet{ Washinoue2023}.

Other assumptions are related to the heating term. The main assumption here is that $\bar{W}_\textrm{k}$ should obey a scaling law with respect to the local density value (or equivalently temperature since pressure is assumed to be constant). Specifically, we assume that
\begin{equation}
    \bar{W}_\textrm{k} \propto \rho^x \quad \Leftrightarrow \quad \bar{W}_\textrm{k} \propto T^{-x}, \label{eq:locality_asummption}
\end{equation}
where $x$ is a free parameter. The validity of this assumption and the resulting value (or range of values) that $x$ is supposed to take are investigated in Sect.~\ref{sect:res_locality}. From now on, we refer to this assumption shortly as the locality assumption, as it explicitly connects $\bar{W}_\textrm{k}$ to the local plasma properties in the loop. We note that this is not strictly justified from the physical point of view, since $\bar{W}_\textrm{k}$ is inherently the property of the kink waves in the coronal loop carrying energy from below the corona, and so its properties should depend strongly on the plasma properties below the corona, including the characteristics of possible wave drivers.

The final assumption is that $\zeta$ and $L_{\perp, \textrm{VD}}$ vary only slowly along the loop, or at least that their variation is negligible compared to the temperature gradient, so that they can be treated as constants in the derivations. In Sect.~\ref{sect:methods_numerics} we use prescriptions for $\zeta$ and $L_{\perp, \textrm{VD}}$ based on the works of \citet{VanDoorsselaere2025} and \citet{McMurdo2026} and find that even for the largest loops considered here, the values for $\zeta$ and $L_{\perp, \textrm{VD}}$ change along the loop by at most 1\% and 5\%, respectively, which is much smaller than the temperature change of around 2 orders of magnitude.

In \citet{Martens2010}, the temperature profile and scaling laws come as a result of solving a second-order ordinary differential equation in terms of $\eta = (T/T_\textrm{max})^{7/2}$ \citep[Eq.~(9) in][rewritten using the notation in this work and the exact form of assumed kink wave heating]{Martens2010}:
\begin{equation}
    \epsilon \eta'' = \eta^{(2\alpha-4)/7} - \xi\eta^{(1-3x)/7}  \label{eq:Martens_original}
\end{equation}
Coefficients $\epsilon$ and $\xi$ are actually expressions that depend on loop parameters and physical constants \citep[Eqs.~(5) and (6) in][]{Martens2010}, which arise from rewriting the energy equation (Eq.~\eqref{eq:energy_equilibrium}) in the form shown in Eq.~\eqref{eq:Martens_original}. Looking at Eq.~\eqref{eq:Martens_original} alone, these two coefficients can be viewed as free parameters, whose values are determined by setting two additional boundary conditions, meaning that a total of four boundary conditions are prescribed:
\begin{align}
    \eta(0)&=0,\\
    \eta(L)&=1,\\
    \eta'(0) = \eta'(L) &= 0.
\end{align}
Thus, solving Eq.~\eqref{eq:Martens_original} with the above boundary conditions yields the temperature profile along the loop and determines the values of coefficients $\epsilon$ and $\xi$, which in turn directly yields the two scaling laws for $T_\textrm{max}$ and $E_\textrm{H}$. 

Hence, the temperature profile and scaling laws given in \citet{Martens2010}, again rewritten in terms of our notation and assumptions about the main source of heating, are
\begin{align}
    T(s) &= T_\textrm{max}\;\tau(s), \label{eq:T(s)} \\
    T_\textrm{max} &= (c_\textrm{T}PL)^{\Tilde{\alpha}/3}, \label{eq:Tmax_gen} \\
    E_\textrm{H}(s) &= 2\kappa I^2 \left(1-\frac{3x}{8}\right)\,  \tau(s)^{(1-3x)/2}\, c_\textrm{T}^{7\Tilde{\alpha}/6}\, P^{7\Tilde{\alpha}/6}\,L^{7\Tilde{\alpha}/6-2}. \label{eq:EH_gen_Martens}
\end{align}
We introduced several new symbols to simplify the expressions. The scaling exponent $\alpha$ in the expression for radiative cooling is replaced by $\Tilde{\alpha} = 12/(11-2\alpha)$. This simplification is particularly useful in case the scaling laws are applied to solar coronal conditions, as there typically $\alpha=-1/2$ \citep[e.g.][]{Martens2010, Washinoue2023}, which corresponds to $\Tilde{\alpha}=1$. The normalised temperature profile $T(s)/T_\textrm{max}$ is defined as
\begin{equation}
    \tau(s) = \left[ \beta_r^{-1} \left(\frac{s}{L}; \frac{2}{4-(2+x)\Tilde{\alpha}}, \frac{1}{2}\right)\right]^{[\Tilde{\alpha}/3] \cdot [2/(4-(2+x)\Tilde{\alpha})]}, \label{eq:cbeta}
\end{equation}
where $\beta_r^{-1}$ denotes the inverse of the regularised incomplete beta function, which is given by
\begin{equation}
    \beta_r(x; a,b) = \frac{\int_0^x t^{a-1}(1-t)^{b-1}\, dt}{\int_0^1 t^{a-1}(1-t)^{b-1}\, dt}, \quad x\in [0,1],\ a, b>0
\end{equation}
Constant $c_\textrm{T}$ is defined as
\begin{equation}
    c_\textrm{T} = \frac{2}{\nu k_\textrm{B}I}\sqrt{\frac{1-f + f\zeta^2}{(1 - f + f\zeta)^{2}}\frac{\chi}{(2\alpha+3)\kappa}},
\end{equation}
where the index suggests in which scaling law (in this case for $T_\textrm{max}$) it appears for the first time. We also introduced
\begin{align}\label{eq:integral}
    I &= \int_0^1 \frac{t^{(7-2\alpha)/4}}{\sqrt{1-t^{(5-2\alpha-3x)/2}}}\, dt \nonumber \\ 
    &= \frac{2}{5-2\alpha-3x}\ \beta \left(\frac{11-2\alpha}{10-4\alpha-6x}, \frac{1}{2}\right),
\end{align}
where $\beta$ stands for the beta function. 

Equations~\eqref{eq:EH_gen_Martens} and \eqref{eq:integral} provide two restrictions on $x$. On one hand, the heating rate has to be positive, meaning $1-3x/8>0$. On the other hand, arguments of the beta function in Eq.~\eqref{eq:integral} have to be positive, which translates to $10-4\alpha-6x>0$, since $\alpha$ is certainly smaller than $11/2$. Together, this results in the following upper limit for $x$: 
\begin{equation}\label{eq:integral_condition}
    x < \frac{\min(5-2\alpha, 8)}{3}.
\end{equation}

From Eqs.~\eqref{eq:EH_init} and \eqref{eq:EH_gen_Martens} it is straightforward to derive the scaling law for $\bar{W}_\textrm{k}$:
\begin{equation}
    \bar{W}_\textrm{k}(s) = \left(1-\frac{3x}{8}\right)^{2/3}\tau(s)^{-x}\; c_\textrm{W}^{2/3}\,P^{(2\Tilde{\alpha}+1)/3}\,L^{(2\Tilde{\alpha}-4)/3}, \label{eq:Wk_gen_Martens}
\end{equation}
where $c_\textrm{W}$ is another constant:
\begin{equation}
    c_\textrm{W} = c_\textrm{T}^{\Tilde{\alpha}} \cdot 2\kappa I^2\, L_{\perp, \textrm{VD}}\sqrt{\frac{\mu m_\textrm{p}}{(1-f+f\zeta) \nu k_\textrm{B}}}.
\end{equation}

\subsection{Numerical RTVS-based approach} \label{sect:methods_numerics}

In the next step, we move away from the analytical approach to relax some of the assumptions needed in Sect.~\ref{sect:methods_analytics}, while retaining the setting of a static and symmetric loop. First, we move away from the assumption that $\bar{W}_\textrm{k}$ scales with temperature. Instead, we use the kink wave energy equations from \citet{VanDoorsselaere2025} and \citet{McMurdo2026} (without the kink wave reflection term) to model the distribution of both propagating components $W_\textrm{k}^{\pm}$ inside the loop:
\begin{equation}
\frac{\partial W_\mathrm{k}^\pm}{\partial t}
+ \nabla \cdot (\vec{Q}_0^\mp W_\mathrm{k}^\pm)
+ \frac{W_\mathrm{k}^\pm}{2} \nabla \cdot \vec{\varv}
= - \frac{1}{L_{\perp, \textrm{VD}}}
\frac{1}{\sqrt{\rho_\mathrm{e}}}
(W_\mathrm{k}^\pm)^{3/2}.
\end{equation}
In the 1D static case, which we are investigating, these equations simplify to
\begin{equation}\label{eq:numeric_Wk}
    (Q^\mp W_\textrm{k}^\pm)' = -E_\textrm{H}^\pm.
\end{equation}
We introduce
\begin{equation}
    Q^\pm = \pm \alpha_\textrm{k} B \label{eq:Q_1D}
\end{equation}
as the 1D static form of Q-variables \citep{VanDoorsselaere2024}, where $B$ is the background static magnetic field and
\begin{equation}
    \alpha_\textrm{k} = \sqrt{\frac{2}{\mu_0(\rho_\textrm{i} + \rho_\textrm{e})}} = \sqrt{\frac{2}{\mu_0} \frac{\nu k_\textrm{B}}{\mu m_\textrm{p}} \frac{(1 - f + f\zeta)^{2}}{\zeta} \frac{T}{P}}. \label{eq:alpha_kink}
\end{equation}
Magnetic permeability is denoted by $\mu_0$. The parameter $\alpha_\textrm{k}$ carries the meaning of the ratio between the kink wave phase speed and the magnetic field strength. We do not include possible reflection of kink waves in our analysis.

We also drop the assumption that pressure is constant inside the loop and instead use the equation of hydrostatic equilibrium \citep[as in e.g.][]{Serio1981}:
\begin{equation}\label{eq:numeric_P}
    P' = -\frac{\mu m_\textrm{p}}{\nu k_\textrm{B}} \frac{P}{T} g,
\end{equation}
where gravitational acceleration $g$ is given by
\begin{equation}\label{eq:numeric_g}
    g = g_0 \cos\left(\frac{\pi}{2} \frac{s}{L}\right),
\end{equation}
with $g_0=274$ m s$^{-2}$ being equal to the gravitational acceleration at the solar surface.

Finally, the energy equation~\eqref{eq:energy_equilibrium} can be rewritten as
\begin{equation}
    E_H - \frac{1-f + f\zeta^2}{(1 - f + f\zeta)^{2}}\frac{\chi}{\left(\nu k_\textrm{B}\right)^2} P^2 T^{\alpha-2} + (\kappa T^{5/2} T')' = 0
\end{equation}
using Eqs.~\eqref{eq:Fc}, \eqref{eq:n_definition}, \eqref{eq:ER}, \eqref{eq:ideal_gas_law} and \eqref{eq:Lambda}. By expressing $T''$ from the equation above explicitly, we derive the following differential equation for temperature:
\begin{equation}\label{eq:numeric_T}
    T'' = \frac{1}{\kappa} \left(\frac{1-f + f\zeta^2}{(1 - f + f\zeta)^{2}}\frac{\chi}{\left(\nu k_\textrm{B}\right)^2} P^2 T^{\alpha-2} - E_\textrm{H} \right) T^{-5/2} - \frac{5}{2}T^{-1}(T')^2.
\end{equation}
We set $\alpha=-1/2$ as that corresponds to the solar corona, as we did in Sect.~\ref{sect:methods_analytics}. Note that $E_\textrm{H}$ can also be explicitly written in terms of pressure and temperature using Eq.~\eqref{eq:EH_total}.

Equations~\eqref{eq:numeric_Wk}, \eqref{eq:numeric_P} and \eqref{eq:numeric_T} form a system of ordinary differential equations with the unknowns $W_\textrm{k}^\pm, P$ and $T$, which can be solved numerically. The numerical code solving this system is publicly available\footnote{\url{https://github.com/lukabano/RTVS_loop_kink}}. In this code, the system of equations is solved in terms of variables $\Tilde{P}=\ln (P/P_0), \Tilde{T} = T^{7/2}$ and $\Tilde{S}^\pm = (\mp Q^\mp W_\textrm{k}^\pm)^{-1/2}$, since this results in the simplification of the system and improved numerical stability. The resulting system of equations is
\begin{align}
    \Tilde{P}' &= -\frac{\mu m_\textrm{p}}{\nu k_\textrm{B}} g_0 \cos\left(\frac{\pi}{2} \frac{s}{L}\right) \Tilde{T}^{-2/7}, \label{eq:numeric_P_transformed}\\
    \Tilde{T}'' &= \frac{1-f + f\zeta^2}{(1 - f + f\zeta)^{2}} \frac{7\chi}{2\kappa \left(\nu k_\textrm{B}\right)^2} P_0^2 e^{2\Tilde{P}} \Tilde{T}^{(2\alpha-4)/7} - \frac{7}{2\kappa L_{\perp, \textrm{VD}}} \cdot\nonumber \\ & \sqrt{\frac{(1-f + f\zeta)\Tilde{T}^{2/7}}{ P_0e^{\Tilde{P}}}} \frac{1}{(\alpha_\textrm{k} B)^{3/2}} \left(\left(\Tilde{S}^+\right)^{-3/4} + \left(\Tilde{S}^-\right)^{-3/4}\right) \label{eq:numeric_T_transformed}\\
    \left(\Tilde{S}^\pm\right)' &= \mp \frac{\left(\mu_0\zeta\right)^{3/4}}{2^{7/4}L_{\perp, \textrm{VD}}\, B^{3/2}(1-f + f\zeta)} \left(\frac{\mu m_\textrm{p} P_0 e^{\Tilde{P}}}{\nu k_\textrm{B}\Tilde{T}^{2/7}}\right)^{1/4} \label{eq:numeric_Wk_transformed}
\end{align}

The assumptions that $\zeta$ and $L_{\perp, \textrm{VD}}$ are approximately constant can also be relaxed by including the expansion of the loop with height, consistent with keeping the pressure balance between the loop and the external plasma. Hence, we use the following prescriptions for $B$, $R$ and $\zeta$, adjusted appropriately from \citet{VanDoorsselaere2025} to the case of a coronal loop:
\begin{align}
    B(s) &= B_0 \left(1+\frac{2L}{R_\sun\pi}\,\sin\left(\frac{\pi s}{2L}\right)\right)^{-2}, \label{eq:background_B} \\
    R(s) &= R_0 \sqrt{\frac{B_0}{B(s)}}, \label{eq:background_R}\\
    \zeta(s) &= (\zeta_0 - 1)\exp\left(-\frac{2L}{5 R_\sun \pi}\,\sin\left(\frac{\pi s}{2L}\right)\right) + 1. \label{eq:background_zeta}
\end{align}

Five boundary conditions are required to solve this boundary value problem. The pressure at the bottom of the loop is fixed to a set value $P_0$. $T(0)=T_0 = 2\cdot 10^4$~K is set as a point where we can reasonably prescribe $T'(0)=0$ as in \citet{Martens2010} due to low temperature gradients in the chromosphere and the fact that the effect on the solution compared to choosing a small positive value for $T'(0)$ is minimal. The maximum temperature at the top is ensured by setting $T'(L)=0$. Finally, we require the wave energy density components to be symmetric, which is obtained by setting $W_\textrm{k}^+(L) = W_\textrm{k}^-(L)$.

Given that this approach builds upon the basic idea of \citet{Serio1981} and that the scaling laws they obtained are sometimes referred to as the RTVS scaling laws, we find it fitting to refer to this method as the RTVS-based method in the rest of this paper.

\subsection{UAWSoM} \label{sect:methods_UAWSoM}

In this work, we also make use of the \mbox{UAWSoM} model, introduced in \citet{VanDoorsselaere2025} and  \citet{McMurdo2026}, where all the details about the mathematical and physical background and initial results can be found. The main points are repeated here for the sake of completeness, with several adjustments made to model a coronal loop in 1D.

\begin{table*}[ht]
    \caption{Summary of the three methods used}
    \centering
    \begin{tabular} { l | l  l  l }
        \hline \hline
        method & analytical RTV-based & numerical RTVS-based & UAWSoM\\
        \hline
        time dependence & no & no & yes\\
        static case only & yes & yes & no\\
        loop leg symmetry & yes & yes & not forced\\
        pressure treatment & constant pressure & hydrostatic equilibrium & full (M)HD treatment \\
        $W_\mathrm{k}^\pm$ treatment & $\bar{W}_\mathrm{k,tot} \propto T^{-x}$ & UAWSoM eqs. & UAWSoM eqs. \\
        no. of equations & 1 (energy cons.) & 4 (for $P, T$ and $W_\mathrm{k}^\pm$) & 5 (for $P, \varv, E$ and $W_\mathrm{k}^\pm$)\\
        parameters specified at the footpoints & - & $P, T$ & "left": $W_\mathrm{k}^-$; "right": $W_\mathrm{k}^+$\\
        \hline
    \end{tabular}
    \label{tab:methods_summary}
\end{table*}

The governing equations in the \mbox{UAWSoM} model are based on the MHD equations, with additional terms for Alfvén and kink wave heating and wave pressure. As the focus of this work is on the kink wave heating, the Alfvén wave heating terms are neglected. Hence, we consider the following simpler system of equations in 1D along the loop:
\begin{align}
\frac{\partial \rho}{\partial t} + \frac{\partial (\rho \varv)}{\partial s} = 0,&
\label{eq:UAWSoM_continuity}\\
\frac{\partial (\rho \varv)}{\partial t}
+ \frac{\partial}{\partial s} \left( \rho \varv^2
 + P - \frac{B^2}{2\mu_0}
 + P_\mathrm{k} \right)
= -\rho g,&  \label{eq:UAWSoM_momentum} \\
\frac{\partial}{\partial t} \left(
\rho \frac{\varv^2}{2}
+ \frac{P}{\gamma - 1}
+ \frac{B^2}{2\mu_0}
+ W_\mathrm{k}^+ + W_\mathrm{k}^-
\right) & \nonumber \\ 
+ \frac{\partial}{\partial s} \left(
\left[
\rho \frac{\varv^2}{2}
+ \frac{P}{\gamma - 1}
- \frac{B^2}{2\mu_0} + P_\mathrm{k}
\right] \varv
+ Q^- W_\mathrm{k}^+
+ Q^+ W_\mathrm{k}^-
\right) & \nonumber \\ 
= -E_\textrm{R} - \nabla\cdot F_\textrm{c}
- \rho \varv g
+  \frac{\zeta - 1}{\zeta + 1} &
P_\mathrm{k} \frac{\partial \varv}{\partial s},
\label{eq:UAWSoM_energy} \\
\frac{\partial W_\mathrm{k}^\pm}{\partial t}
+ \frac{\partial}{\partial s} \left(Q^\mp W_\mathrm{k}^\pm\right)
+ \frac{W_\mathrm{k}^\pm}{2} \frac{\partial \varv}{\partial s}
= - E_\textrm{H}^\pm&.
\label{eq:UAWSoM_Wk}
\end{align}
The induction equation simplifies to $B$ being constant in time. Q-variables are now used in their complete form:
\begin{equation}
    Q^\pm = \varv\pm \alpha_\textrm{k} B
\end{equation}Gravity, magnetic field, loop radius and density contrast profiles are the same as before (Eqs.~\eqref{eq:numeric_g}, (\ref{eq:background_B}-\ref{eq:background_zeta})). The polytropic index is denoted by $\gamma$ and is set to $5/3$, corresponding to an ideal monoatomic gas. We also set $A_\textrm{He}=0.1$ in all \mbox{UAWSoM} simulations. Kink wave pressure $P_{\mathrm{k}}$ is equal to
\begin{equation}
P_\mathrm{k}=\frac{W^+_\mathrm{k}+W^-_\mathrm{k}}{2\mu_0\rho\alpha_\textrm{k}^2}.
\label{eq:P_k}
\end{equation}

Equations (\ref{eq:UAWSoM_continuity}-\ref{eq:UAWSoM_Wk}) are solved within the framework of \texttt{MPI-AMRVAC} \citep{Keppens2003, Keppens2023}. Notably, the magnetic field splitting technique is used to improve the numerical stability and accuracy \citep{Tanaka1994, Xia2018}.

The differences between the set of equations considered in the previous section (Eqs.~\eqref{eq:numeric_Wk}, \eqref{eq:numeric_P} and \eqref{eq:numeric_T}) and the \mbox{UAWSoM} model can be summarised as follows. In \mbox{UAWSoM} we abandon the assumptions of the loop being static and in hydrostatic equilibrium. Magnetic and kink wave pressure terms are also no longer neglected. Lastly, the power-law radiative cooling approximation is replaced with the cooling curve by \citet{Colgan2008}.

The boundary conditions are set at $s=0$ and $s=2L$, i.e.\ the boundaries are located at loop footpoints. Hence, they are very similar for both boundaries and are equivalent to the ones used in \citet{McMurdo2026} for the bottom boundary. Their formulation results primarily from the assumption that the plasma beyond the domain is in hydrostatic equilibrium, though this does not force the plasma inside the domain to stay in hydrostatic equilibrium, clearly demonstrated by the results in Sect.~\ref{sect:res_uawsom}. When written in mathematical notation, the boundary conditions are
\begin{align}
    \diff{\rho}{s} &= \frac{g\rho^2}{P},\\
    \diff{P}{s} &= -\rho g,\\
    \diff{\varv}{s} &= 0, \\
    W_\mathrm{k}^- &= W_\mathrm{k0}, \\
    \diff{W_\mathrm{k}^+}{s} &= 0.
\end{align}
for the "left" boundary, with the last two being replaced by
\begin{align}
    W_\mathrm{k}^+ &= W_\mathrm{k0}, \\
    \diff{W_\mathrm{k}^-}{s} &= 0,
\end{align}
at the other ("right") boundary. Here $W_\mathrm{k0}$ stands for the kink wave energy density at the loop footpoint. With this setup, kink wave energy density at the loop base is fixed in time, while density, pressure and velocity at the boundaries are allowed to adjust to the varying conditions inside the loop domain.

The loop is initially assumed to be isothermal with $T=1$~MK and stationary, resulting in the initial density being equal to
\begin{equation}
    \rho(t=0) = \rho_0 \exp\left(-\frac{L}{s_\textrm{H}} \frac{1}{\pi} \sin\left(\frac{\pi s}{L}\right)\right),
\end{equation}
where
\begin{equation}
    s_\textrm{H} = \frac{k_\textrm{B} g_0}{m_\textrm{p} T}
\end{equation}
is the isothermal pressure scale height and $\rho_0$ is the footpoint density.
Both kink wave energy components are set symmetrically using the following profile:
\begin{equation}
    W^\pm_\mathrm{k} (t=0) =
    \begin{cases}
        0.4 W_\mathrm{k0}\left(\cos\left(\dfrac{\pi s}{\delta}\right) + 1.5\right), & \text{for } |s-L| > L-\delta, \\
        0.2 W_\mathrm{k0}, & \text{otherwise},
    \end{cases}
\end{equation}
where $\delta = 10$ Mm. This ensures a smooth transition between the boundary and the rest of the domain, as well as a smooth start into the simulation by avoiding that $W^\pm_\mathrm{k} = 0$ anywhere inside the domain.

As a convenience for the reader, the main characteristics and differences between the three presented models are summarised in Table~\ref{tab:methods_summary}. These differences stem primarily from the assumptions behind each of the models and result in e.g. the different modelling capabilities, accuracy of the models and their results (one of the main topics of Sect.~\ref{sect:results}), and boundary conditions for the numerical models.

\section{Results and discussion}\label{sect:results}

\subsection{Results of the analytical model}\label{sect:res_analytic}

The discussion of the results starts by analysing the scaling law expressions obtained in Sect.~\ref{sect:methods_analytics}.
To make the analysis clearer, we focus our attention on the solar atmosphere, for which the radiative losses in the temperature range of our interest can be estimated by $\alpha = -0.5$ and $\chi=10^{-18.8}$ erg cm$^3$ s$^{-1}$  K$^{1/2}$ \citep{Washinoue2023}, the former corresponding to $\Tilde{\alpha}=1$. This also affects Eq.~\eqref{eq:integral_condition}, which simplifies to $x<2$. In this case, we recover the usual exponents from the scaling laws for $T_\textrm{max}$ and $E_\textrm{H}$:
\begin{align}
    T_\textrm{max} &= (c_\textrm{T}PL)^{1/3}, \label{eq:Tmax_Sun} \\
    E_\textrm{H}(s) &= 2\kappa I^2\left(1-\frac{3x}{8}\right) \tau(s)^{(1-3x)/2}\, c_\textrm{T}^{7/6}\, P^{7/6}\,L^{-5/6}. \label{eq:EH_Sun}
\end{align}
In the case of $\bar{W}_\textrm{k}$, the result is
\begin{equation}
    \bar{W}_\textrm{k}(s) = \left(1-\frac{3x}{8}\right)^{2/3}\tau(s)^{-x}\, c_\textrm{W}^{2/3}\,P\,L^{-2/3}, \label{eq:Wk_gen_Martens_Sun}
\end{equation}

\begin{figure*}[ht]
    \centering
    \includegraphics[width=0.33\hsize]{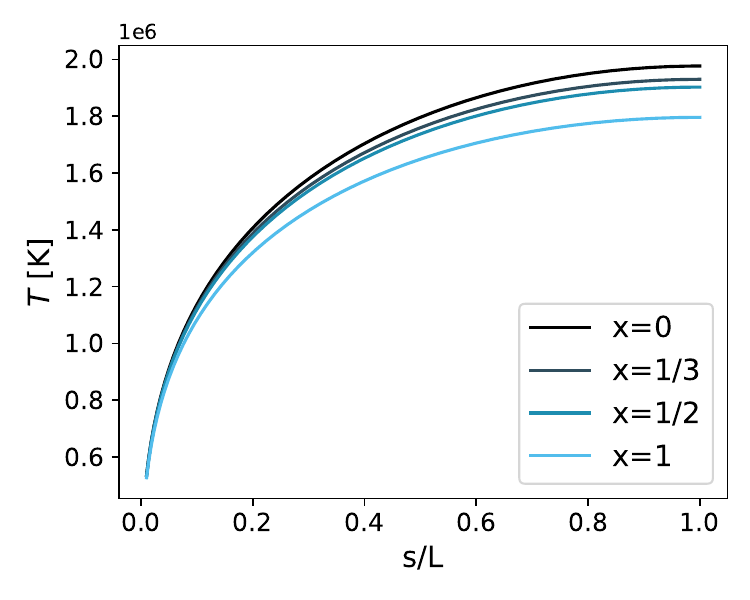}
    \includegraphics[width=0.33\hsize]{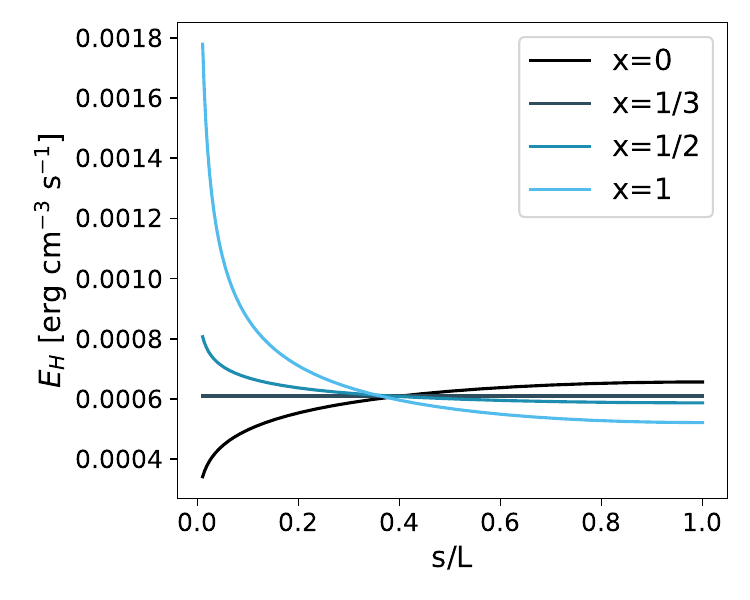}
    \includegraphics[width=0.33\hsize]{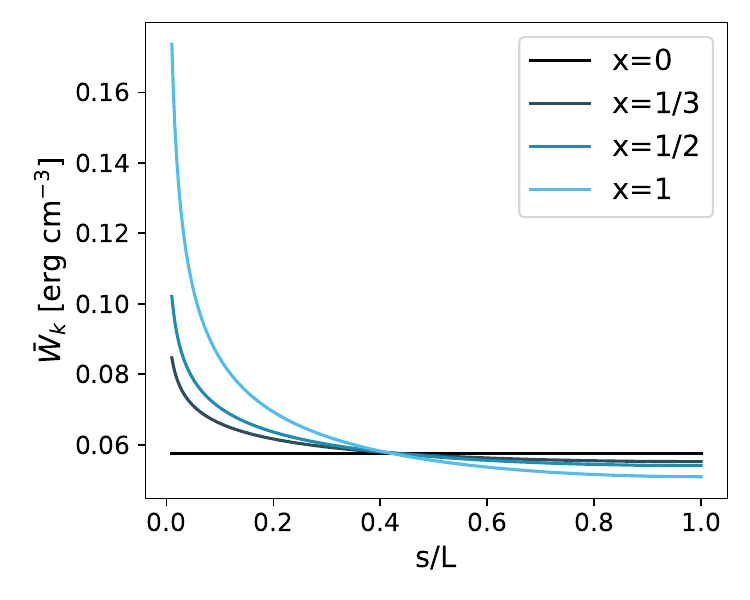}

    \caption{Analytical profiles for $T$ (left panel), $E_\textrm{H}$ (middle panel) and $\bar{W}_\textrm{k}$ (right panel) depending on the value of $x$ for the loop with $P_0=0.5$ erg cm$^{-3}$ and $L=50$ Mm. The profiles are plotted with respect to the normalised loop length, with the bottom 0.5 Mm excluded.
    }
    \label{fig:RTV_plots_analytical} 
\end{figure*}

\begin{figure*}[ht]
    \sidecaption
    \includegraphics[width=12cm]{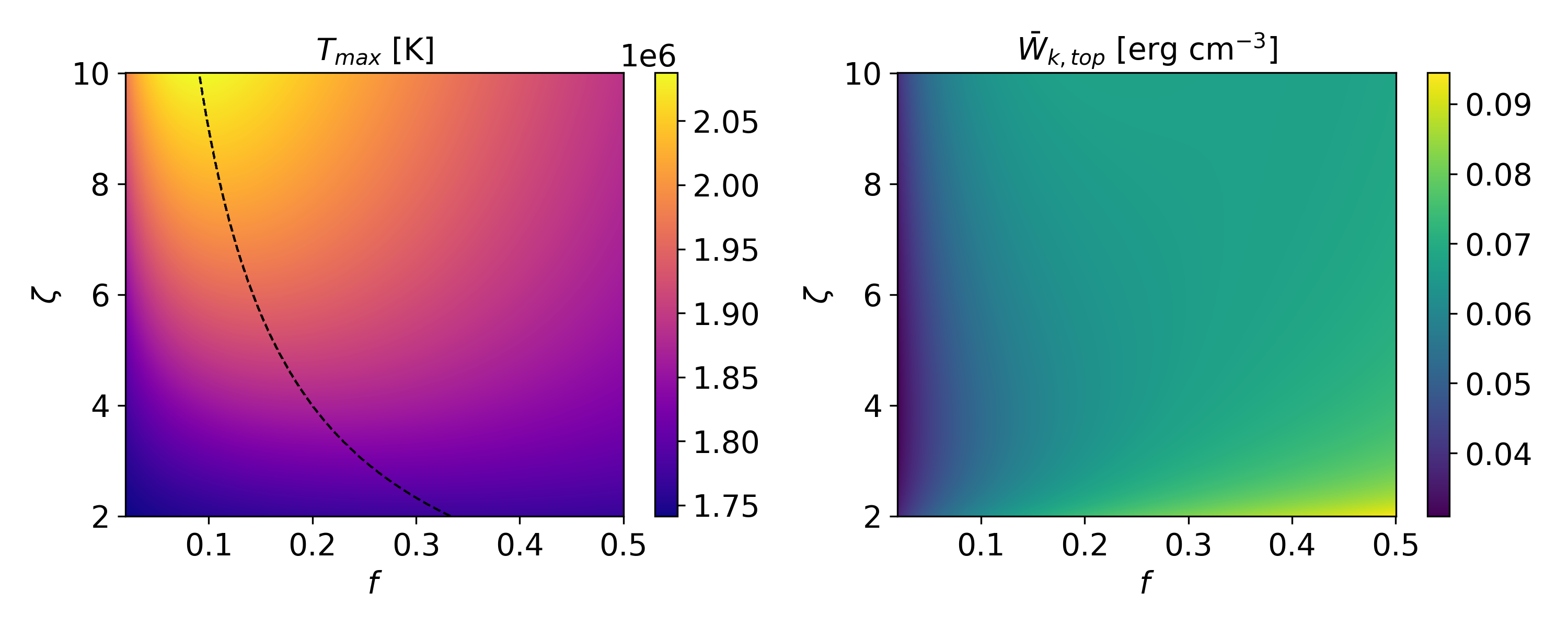}
    \caption{Dependence of $T_\textrm{max}$ (left panel) and $\bar{W}_\textrm{k}$ at the top of the loop (right panel) on the filling factor $f$ and density contrast $\zeta$. The dashed black line in the left panel denotes the value of $f$ for which the maximum value of $T_\textrm{max}$ is attained, for a given $\zeta$.}
    \label{fig:param_study_analytical} 
\end{figure*}

From the theoretical standpoint, there are several interesting cases regarding the value of $x$. Case $x=0$ corresponds to $\bar{W}_\textrm{k}$ being independent of temperature (Eq.~\eqref{eq:locality_asummption}). From Eq.~\eqref{eq:EH_init} it follows that the case $x=1/3$ corresponds to the heating rate $E_\textrm{H}$ being independent of temperature. Finally, case $x=1$ corresponds to the velocity amplitude of the wave $V$ being independent of temperature, which is evident from Eqs.~(43) and (52) in \citet{VanDoorsselaere2025}, the relevant part of which reads
\begin{equation}\label{eq:Wk_rho_Y}
    \bar{W}_\textrm{k} \propto \rho V^2.
\end{equation} 

In this section, we choose $\zeta=5, f=0.1$ and $R=10^8$~cm as the standard loop parameters. We also use the standard value for Spitzer conductivity of $\kappa = 1.7\cdot 10^{-6}$ erg cm$^{-1}$ s$^{-1}$ K$^{-7/2}$ \citep[e.g.][]{Bradshaw2019}. The value $A_{\textrm{He}} = 0$ is chosen as this is typically assumed in similar analytical studies. With these values, the three scaling laws further simplify to
\begin{align}
    T_\textrm{max} &= 1.27\cdot 10^3\cdot I^{-1/3}\cdot &&(PL)^{1/3}, \label{eq:Tmax_simple}\\
    E_\textrm{H}(s) &= 2.50\cdot 10^5\cdot I^{5/6}\, \left(1-\frac{3x}{8}\right)\, \tau(s)^{(1-3x)/2}\cdot \hspace{-5mm} &&P^{7/6}\,L^{-5/6}, \label{eq:EH_simple}\\
    \bar{W}_\textrm{k}(s) &= 4.41\cdot 10^5\cdot I^{2/3}\, \left(1-\frac{3x}{8}\right)^{2/3}\, \tau(s)^{-x} \cdot &&P\,L^{-2/3}. \label{eq:Wk_simple}
\end{align}
All constant coefficients above are in cgs units. The fact that kink wave heating is used to arrive at Eqs.~\eqref{eq:Tmax_simple} and ~\eqref{eq:EH_simple} is reflected in all the coefficients and terms listed before the pressure and loop length, while the power law coefficients remain unaffected. The choice of $x$ evidently affects the final form of the scaling laws and can also be hand-picked to reproduce exactly the first two laws as usually listed \citep{Rosner1978, Bourdin2016, Washinoue2023}: for example, for $x=0.5$ Eq.~\eqref{eq:Tmax_simple} we retrieve the famous expression $T_\textrm{max} = 1.40\cdot 10^3\cdot (PL)^{1/3}$. True novelty lies in Eq.~\eqref{eq:Wk_simple}, relating kink wave energy density to pressure and loop length.

The distribution of temperature, heating rate and kink wave energy density along a coronal loop is also shown graphically in Figure~\ref{fig:RTV_plots_analytical} for a specific case of $P_0=0.5$ erg cm$^{-3}$ and $L=50$ Mm. Distributions are shown for four different values of the parameter $x$; along with the three aforementioned cases a fourth one, $x=0.5$, is also considered, which is chosen based on the results of the numerical RTVS-based approach (see Sect.~\ref{sect:res_locality}). 
Overall, Fig.~\ref{fig:RTV_plots_analytical} shows the difference in the scaling laws resulting from the variation of the parameter $x$. Panels showing the temperature and heating rate profiles are similar in content to Fig.~3 in \citet{Martens2010}, since varying our parameter $x$ has the same role as varying the parameter denoted by $\alpha$ in \citet{Martens2010} (with the meaning of $E_\textrm{H} \propto T^\alpha$; the conversion between the two is $\alpha \rightarrow (1-3x)/2$). The main differences to Fig.~3 of \citet{Martens2010} are in the treatment of radiative losses and the range of $x$ values which are considered.

The main effects of increasing the value of $x$ are small reductions in the loop temperature at the top, redistribution of the heating from the one with the dominant contribution at the loop top to footpoint-dominated heating, and an increase in the kink wave energy density at the footpoint required to sustain the loop. The distribution of $\bar{W}_\textrm{k}$ in the loop is also strongly affected, as the $\bar{W}_\textrm{k}$ profile becomes steeper with increasing $x$ and results in decreasing $\bar{W}_\textrm{k}$ values at the loop top.

While the effect of varying pressure and loop length is clear from the form of the scaling laws, it is less so in the case of varying parameters related to the loop geometry. The effect of the loop radius is the easiest one to deduce, since the radius is only present in Eq.~\eqref{eq:L_perp}, meaning that it does not affect the scaling laws for $T_\textrm{max}$ and $E_\textrm{H}$. For $\bar{W}_\textrm{k}$, one obtains
\begin{equation}
    \bar{W}_\textrm{k} \propto R^{2/3},
\end{equation}
meaning that more kink wave energy density is needed to sustain a thicker loop, which then requires larger wave amplitudes as well: $V \propto R^{1/3}.$

The effect of the filling factor and density contrast on temperature and $\bar{W}_\textrm{k}$ at the loop apex is simultaneously plotted in Fig.~\ref{fig:param_study_analytical}. Pressure, loop length and radius of the loop are the same as in Fig.~\ref{fig:RTV_plots_analytical} and $x=0.5$. The range of values for $f$ and $\zeta$ is chosen to cover values typically found and used in the literature \citep[][and references therein]{Morton2021, Gao2022, VanDoorsselaere2014}.
The left panel shows that the temperature at the top of the loop increases both with increasing $\zeta$ and $f$ in the shown range, though we note that for each value of $\zeta$ there is a specific value of $f$ for which $T_\textrm{max}$ is largest; this value of $f$ decreases with $\zeta$ and becomes as low as 0.1 when $\zeta$ is as large as 10. Overall, $T_\textrm{max}$ shows a variation of around 20\% in the examined range of values for $\zeta$ and $f$, which is significant and comparable to the variation due to the uncertainty in the appropriate value of $x$ (Fig.~\ref{fig:RTV_plots_analytical}, left panel).
Kink wave energy density in the loop (shown for the top of the loop specifically, but the conclusion is valid everywhere in the loop) depends significantly on $f$ and $\zeta$, though the influence becomes limited as $f>0.2$ and $\zeta > 3$. The predicted values of $\bar{W}_\textrm{k}$ for high $f$ and low $\zeta$ are more than double compared to the case of low $f$ and high $\zeta$.
We also point out that for loops with different values of $P, L$ or $x$ compared to the scenario examined in Fig.~\ref{fig:param_study_analytical} result in a simple rescaling of the values shown in Fig.~\ref{fig:param_study_analytical}, so the effects of $\zeta$ and $f$ remain largely the same.

\subsection{Results of the RTVS-based method}\label{sect:res_numeric}

Next, the results of solving Eqs.~(\ref{eq:numeric_P_transformed}-\ref{eq:numeric_Wk_transformed}) numerically are shown. In the numerical results throughout this paper, we set $\zeta_0=5, f=0.1, R_0=1$ Mm and $B_0=20$ G, unless stated otherwise. The same values are used in \citet{McMurdo2026}. In coronal loops, the magnetic field is often found to be much stronger \citep{Brooks2021}; however, we found that the choice of $B_0$ has a negligible effect on the results. We also set $A_{\textrm{He}} = 0$.

\begin{figure}[ht]
    \centering
    \includegraphics[width=0.45\textwidth]{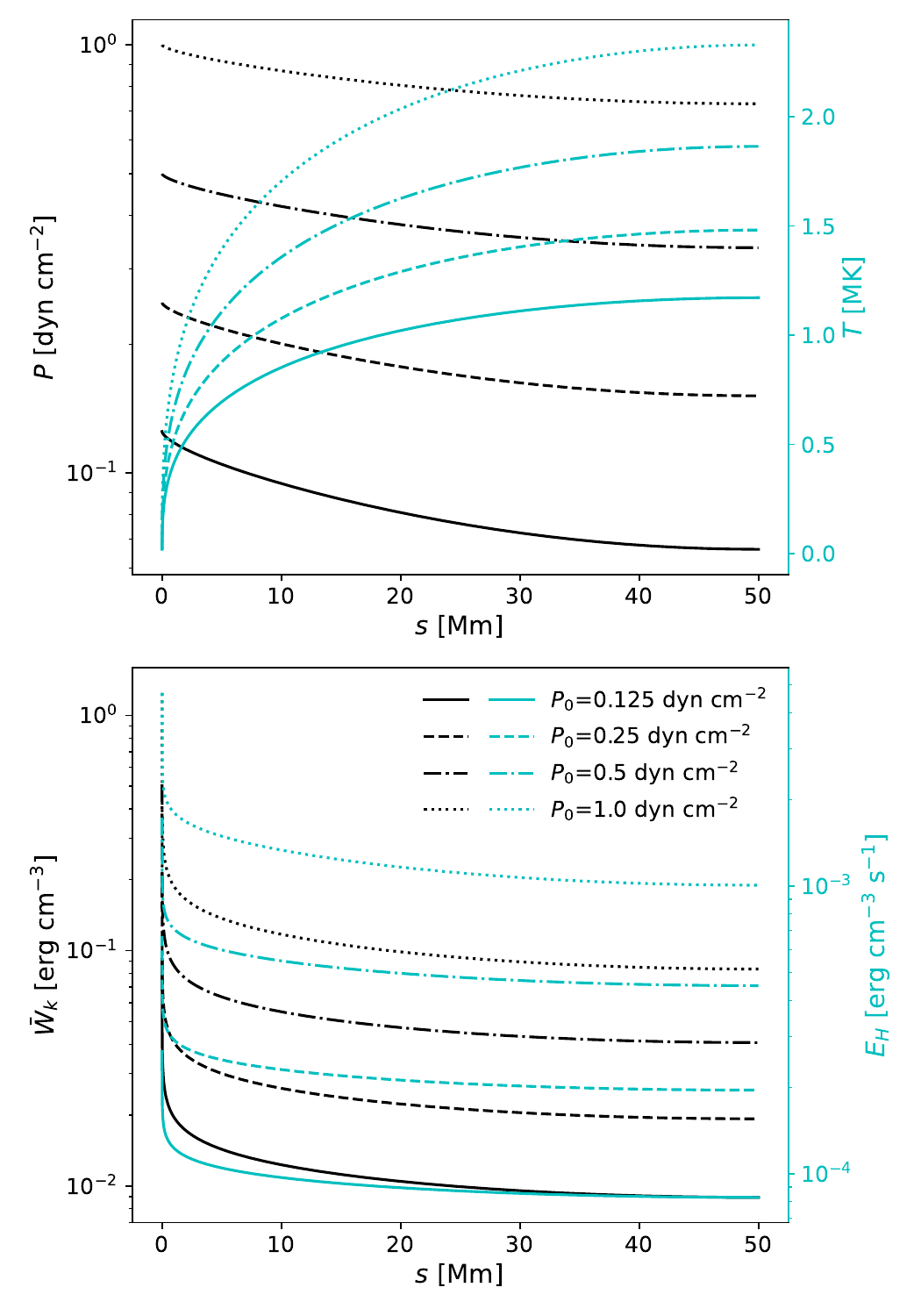}
    \caption{Numerical solutions for $P$ (top panel, black lines), $T$ (top panel, blue lines), $\bar{W}_\textrm{k}$ (bottom panel, black lines) and $E_\textrm{H}$ (bottom panel, blue lines) for a loop with $L=50$ Mm and different values of $P_0$, represented by different line types.}
    \label{fig:numerics_general}
\end{figure}

Fig.~\ref{fig:numerics_general} shows the main characteristics of the solutions to the equations, namely the resulting pressure, temperature, total kink wave energy density and total heating rate. In all four cases $L=50$ Mm, while $P_0$ varies between 0.125 and 1 dyn cm$^{-2}$. The general results are consistent with the analytically obtained scaling laws: an increase in pressure results in higher temperatures, heating rates and kink wave energy densities. We also confirm that for such a large loop the ratio between pressure at the bottom and at the top is substantial, meaning that we expect significant differences to the analytical result (which is investigated in more detail in Sect.~\ref{sect:res_comparison}). Regarding the heating rate, the model predicts that kink wave heating occurs preferentially near the loop footpoints. We also note significant deviations in the heating profile from a straight line in the corresponding log-plot in Fig.~\ref{fig:numerics_general}. This suggests that the heating rate does not follow precisely the exponentially-decaying heating with a single scaling height value used in \citet{Serio1981}, but that this heating scaling height decreases sharply towards the footpoints. We find the heating scale height to vary from roughly one Mm or even less near the footpoints to between several hundred and one thousand Mm near the loop apex, depending on pressure and loop length.

\begin{figure}[ht]
    \centering
    \includegraphics[width=0.4\textwidth]{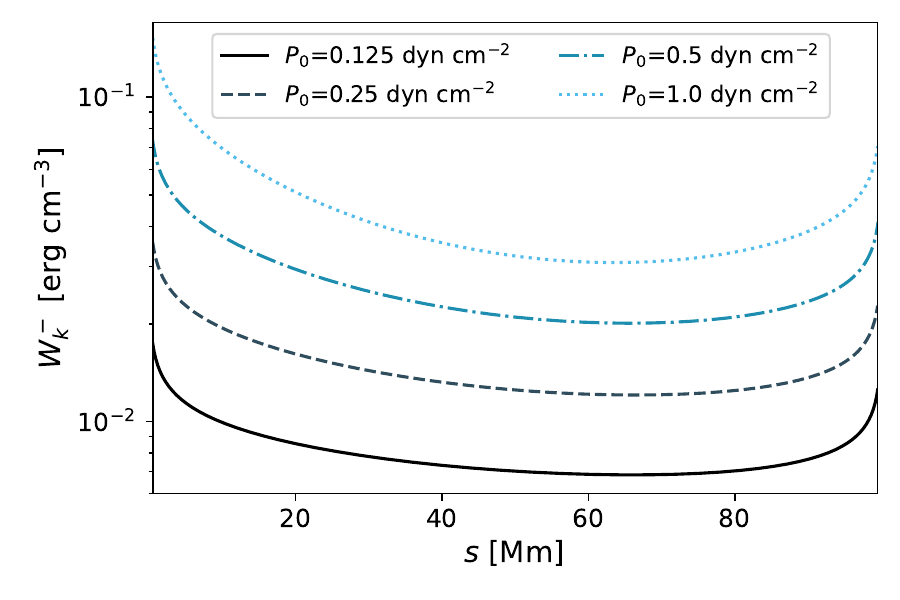}
    \caption{Plot of $W_\textrm{k}^-$ along the whole loop without the 0.5 Mm regions near each of the footpoints. $W_\textrm{k}^-$ for the right half of the loop is determined using the $W_\textrm{k}^+$ results in the left half of the loop due to symmetry.}
    \label{fig:numerics_Wkwhole}
\end{figure}

Since the numerical approach models both propagating kink wave components separately, Fig.~\ref{fig:numerics_Wkwhole} shows one of the components, namely $W_\textrm{k}^-$, along both halves of the loop. $W_\textrm{k}^-$ specifically represents the energy of the waves propagating from the footpoint towards the loop top in our model of one half of the loop. The value of $W_\textrm{k}^-$ in the other half (i.e.\ right half in Fig.~\ref{fig:numerics_Wkwhole}) is equal to $W_\textrm{k}^+$ in the left half, flipped over the loop top, due to the assumed symmetry of the loop. $W_\textrm{k}^-$ profiles are shown for the same four loops considered in Fig.~\ref{fig:numerics_general}. The main thing to note is that the $W_\textrm{k}^\pm$ components themselves are not symmetric, which is a direct result of wave dissipation. The increase of $W_\textrm{k}^-$ towards the footpoint further from the wave source is explained by $W_\textrm{k}$ and $\rho$ being proportional to each other, as noted explicitly in Eq.~\eqref{eq:Wk_rho_Y} for example. In general, the difference in $W_\textrm{k}^\pm$ between the footpoints serves as an indication of how much energy has been dissipated along the loop.

This approach is able to provide coronal loop solutions for a wide range of input pressure and loop length. This may come as a surprise, given that near the footpoint the exponential heating scale height appears to be very low ($<1$ Mm), which would suggest that the loops might become unstable \citep{Serio1981, Aschwanden2002} and could induce thermal non-equilibrium (TNE) cycles \citep{Klimchuk2019, Antolin2022}. Interestingly, the opposite, stabilising effect of the heating scale height being low has also been reported \citep{Pelouze2022}. Nevertheless, the fact that the exponential scale height increases rapidly with height in our case likely plays a crucial role in ensuring the existence of stable solutions. We note, however, that we leave the question of whether an actual loop would be steady or not open to debate for two reasons: the apparent lack of steady-state loop models when using \mbox{UAWSoM} (which may be due to physical causes as well as due to the model limitations, see Sect.~\ref{sect:res_uawsom}) and the fact that the inclusion of the kink wave pressure term in the pressure equation for the RTVS-based model (Eq.~\eqref{eq:numeric_P}), which is automatically included in \mbox{UAWSoM}, resulted in the apparent lack of solutions.

\subsection{Results of UAWSoM}\label{sect:res_uawsom}

Next, we discuss the results of \mbox{UAWSoM} simulations. Two separate aspects are considered. On one hand, we wish to answer the question of how well the assumption that the loop is static holds, which has been used in the other two methods in this work, and what dynamic effects are predicted as a result. On the other hand, this study also serves the important purpose of testing and validating the model in its current state. With this in mind, we aim to provide reasonable predictions, while being aware of all the limitations of our current setup.

\begin{figure}[ht]
    \centering
    \includegraphics[width=0.9\hsize]{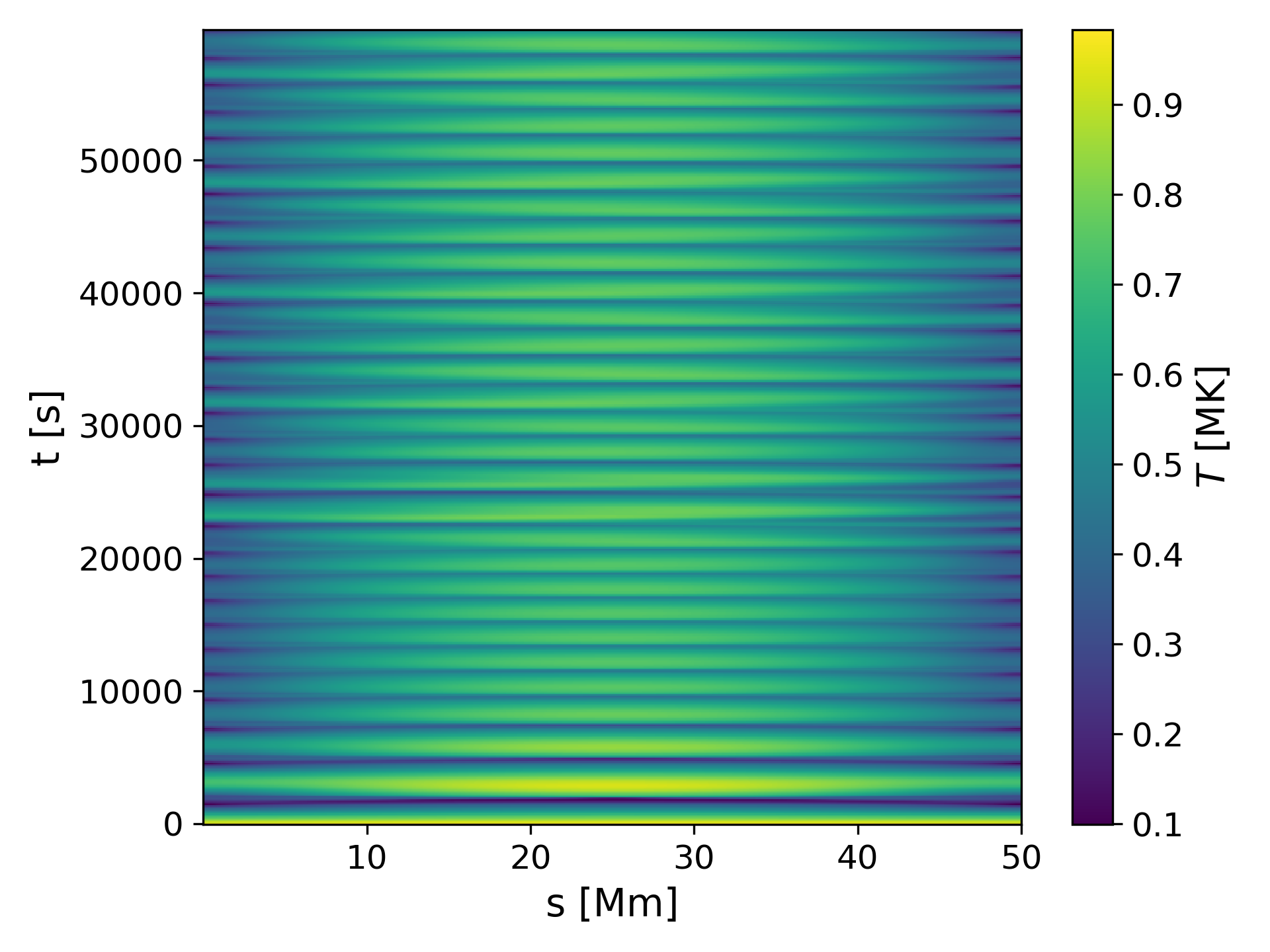}
    \includegraphics[width=0.9\hsize]{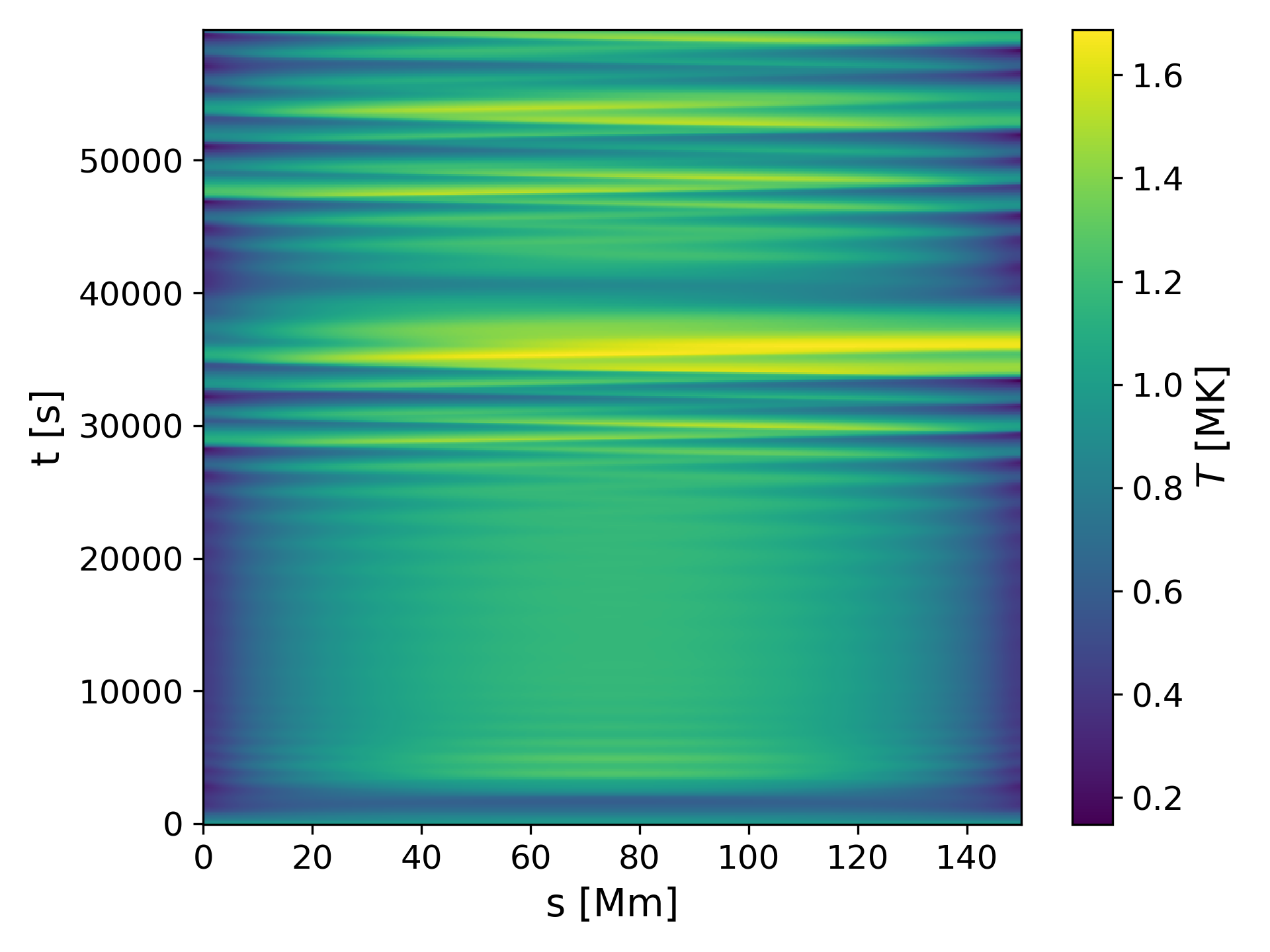}
    \includegraphics[width=0.9\hsize]{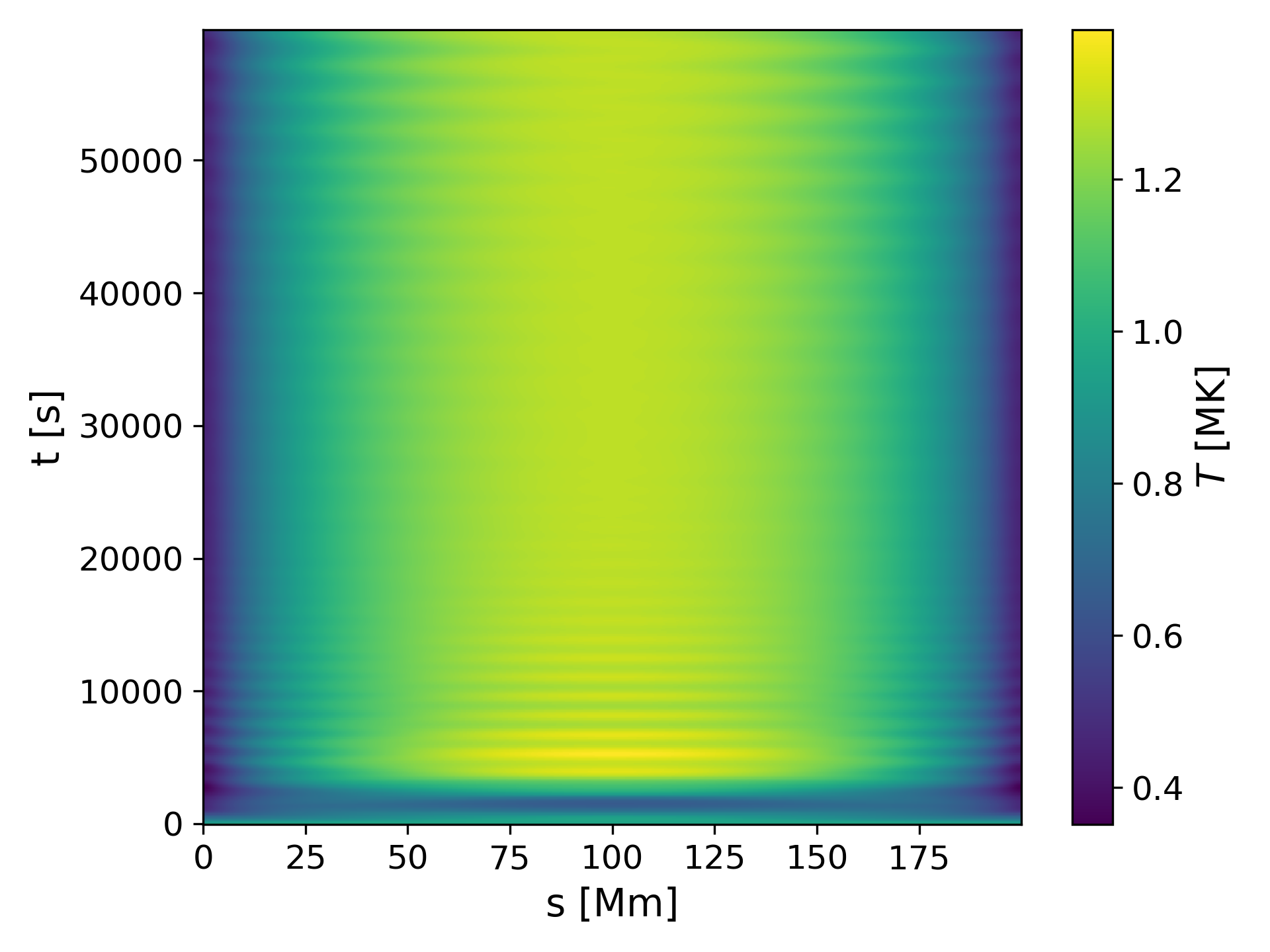}

    \caption{Time evolution of temperature  of the \mbox{UAWSoM} simulations of the loops with lengths $2L=50$ Mm (top), $2L=150$ Mm (middle) and $2L=200$ Mm (bottom). All simulations have the same spatial resolution (with the cell size of 83 km) and use the same `minmod' slope limiter.}
    \label{fig:UAWSoM_T_differentL} 
\end{figure}

\begin{figure*}[ht]
    \sidecaption
    \includegraphics[width=6cm]{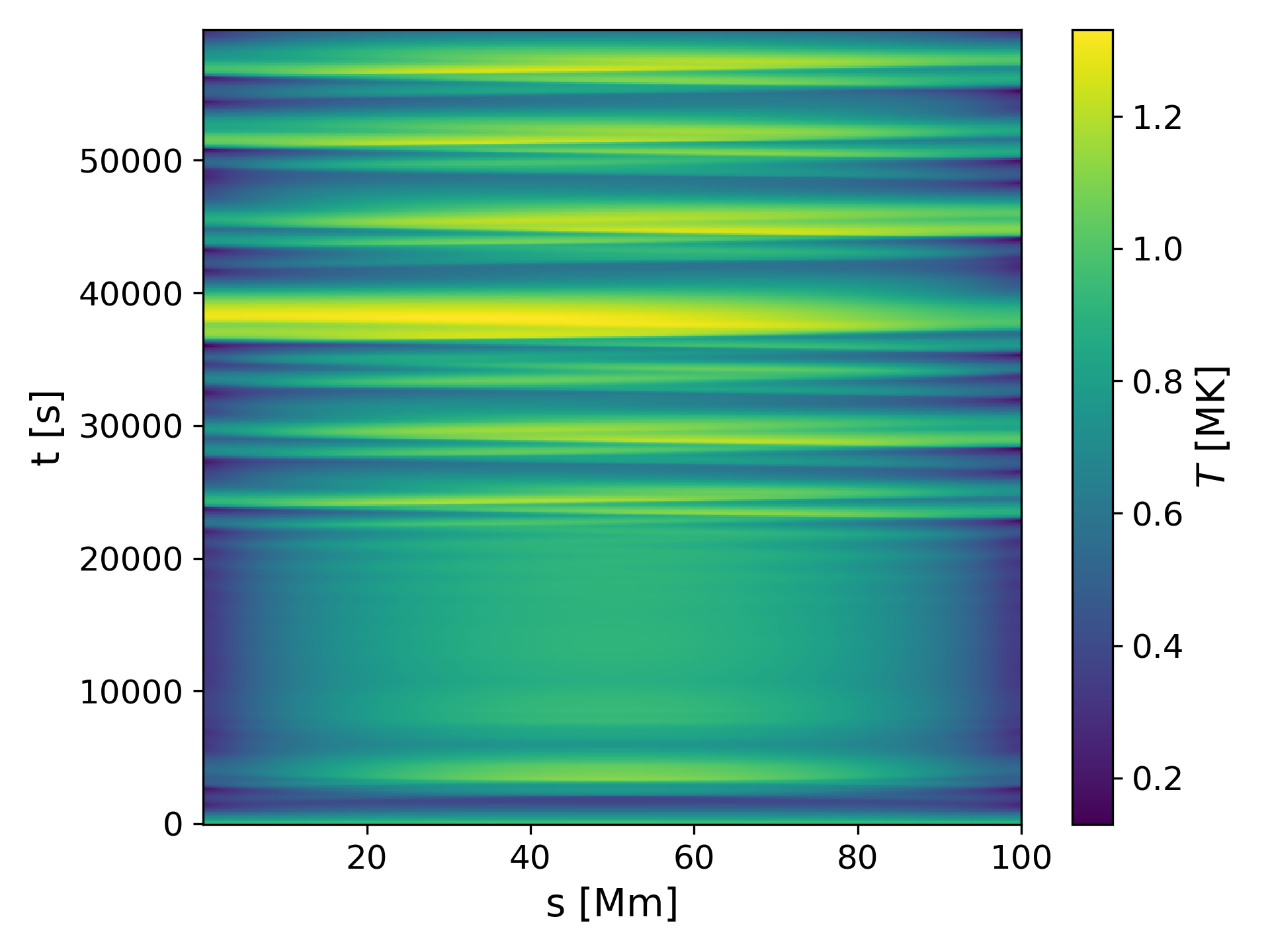}
    \includegraphics[width=6cm]{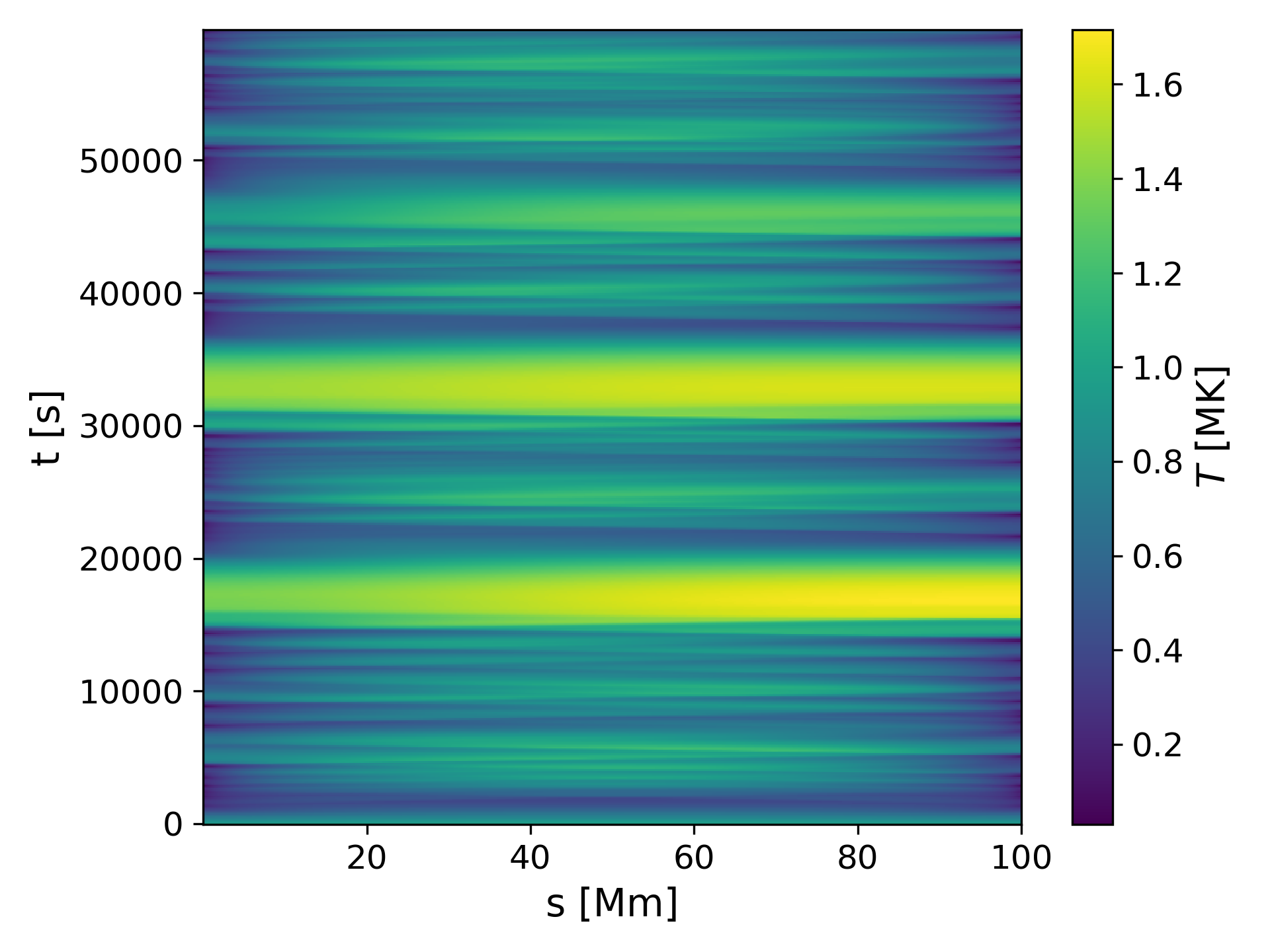}
    \caption{Time evolution of temperature of the \mbox{UAWSoM} simulations of the loops with lengths $2L=100$ Mm and $W_\textrm{k0}=5\cdot 10^{-3}$ erg cm$^{-3}$. The simulation shown in the left panel has a resolution of 1200 cells and uses a diffusive `minmod' limiter, while the one in the right panel has a resolution of 4800 cells and uses a sharper `cada3' limiter.}
    \label{fig:UAWSoM_T_differentres} 
\end{figure*}

Figures~\ref{fig:UAWSoM_T_differentL} and \ref{fig:UAWSoM_T_differentres} depict the variation of coronal loop regimes we encounter in our simulations by showing temperature evolution along the loop over time. Loops of different lengths are shown in Fig.~\ref{fig:UAWSoM_T_differentL}, while effects of using different resolution and slope limiter (`minmod' in the top panel, see e.g. \citet{Toth1996}, and `cada3' in the bottom panel, see \citet{Cada2009}) are shown in Fig.~\ref{fig:UAWSoM_T_differentres}. The amount of wave energy injected in the simulations in Fig.~\ref{fig:UAWSoM_T_differentL} is appropriately adjusted so that temperature and pressure remain similar between different cases. Specifically, for the loop with length $2L=50$ Mm, we set $W_\textrm{k0}=7\cdot 10^{-3}$ erg cm$^{-3}$, while for the other two loops $W_\textrm{k0}$ is adjusted so that it scales with $L^{-2/3}$, which follows from the analytical scaling law in Eq.~\eqref{eq:Wk_gen_Martens_Sun}.

After the first two hours, which is how long it takes before the initial conditions are relaxed, simulations using the diffusive `minmod' limiter enter a quasi-stable phase, with almost no flows present for larger loops (when $2L\gtrsim 100$~Mm) and global oscillations similar to TNE cycles \citep{Froment2018} present for smaller loops. The length of this quasi-stable phase lasts between 5 and 20 hours for the cases we tested, with the duration generally increasing with loop length. Afterwards, an instability grows large enough to trigger emergence and propagation of slow waves. Note that this also eventually happens for the longest of the loops in Fig.~\ref{fig:UAWSoM_T_differentL}, after the time span shown in the figure. Flow velocities can reach up to 70~km~s$^{-1}$ during this stage. Even though the kink wave energy density at the footpoints is set lower for longer than for shorter loops (which results in a lower amount of heating throughout the loops), the fact that this adjustment is based on the footpoint pressure means that the temperature at the loop apex still increases as the loop length is increased due to the increasingly lower densities and pressures near the loop apex.

\begin{figure*}[ht]
    \centering
    \includegraphics[width=0.425\textwidth]{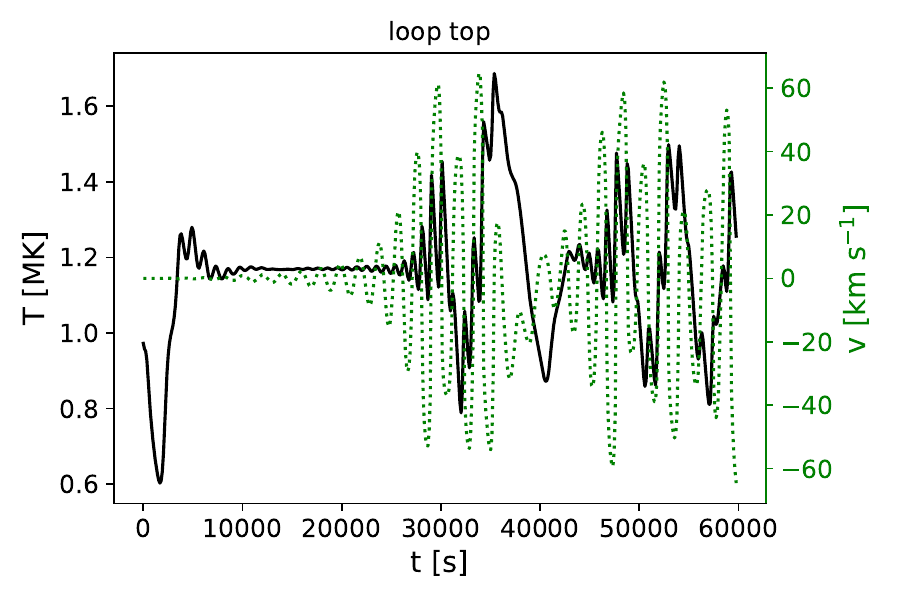}
    \includegraphics[width=0.57\textwidth]{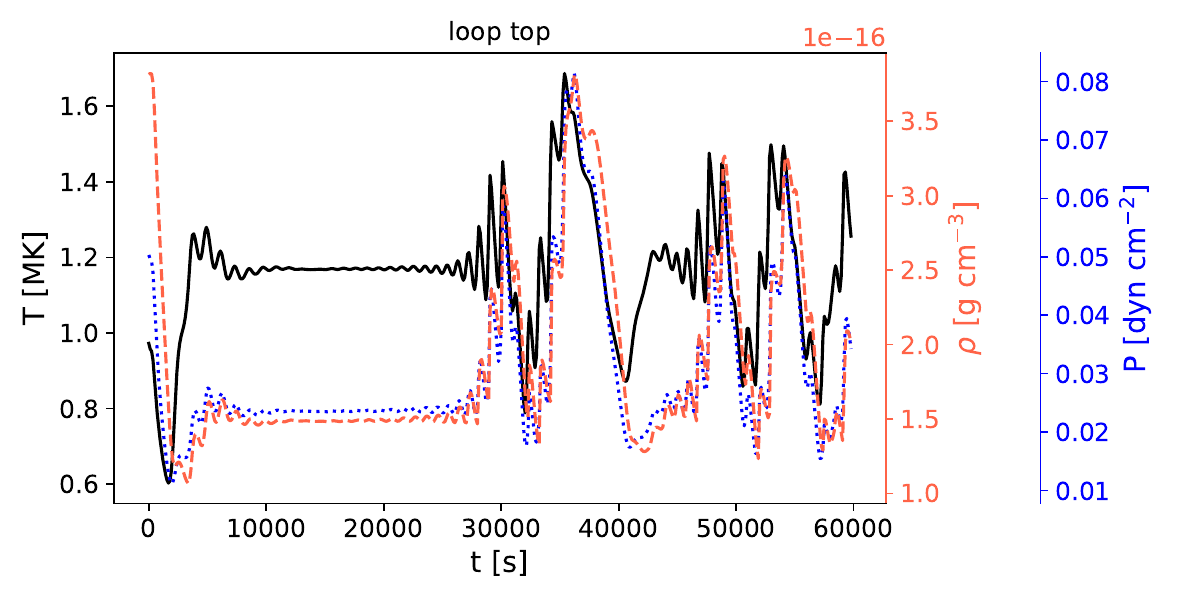}
    \caption{Time evolution of temperature (black solid line in both panels),  velocity (green dotted line; left panel), pressure (blue dotted line; right panel), and density (red dashed line, right panel) at the loop top with length $2L=150$ Mm. The simulation in question is the same one as the one shown in the middle panel in Fig.~\ref{fig:UAWSoM_T_differentL}.}
    \label{fig:UAWSoM_Tv}
\end{figure*}

\begin{figure}[ht]
    \centering
    \includegraphics[width=0.49\textwidth]{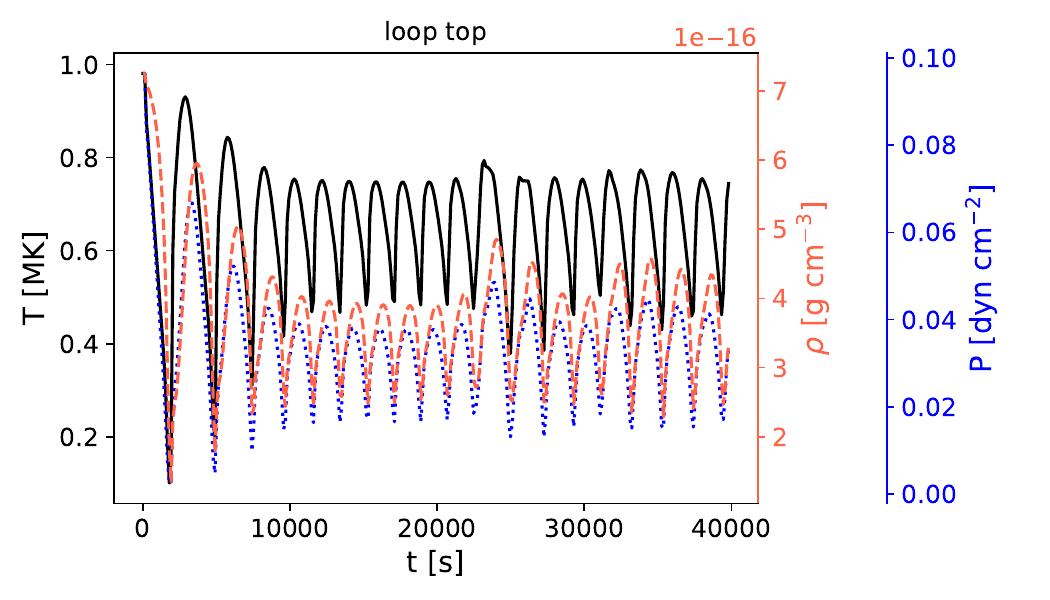}
    \caption{Same as the right panel of Fig.~\ref{fig:UAWSoM_Tv}, but for the loop with length $2L=50$~Mm (top panel, Fig.~\ref{fig:UAWSoM_T_differentL}) and for the first 40000~s only.}
    \label{fig:UAWSoM_Tprho_16k}
\end{figure}

The emergence and development of this instability can also be seen in the left panel of Fig.~\ref{fig:UAWSoM_Tv} for the simulation with the loop length of 150 Mm and `minmod' limiter, the same one shown in the middle panel in Fig.~\ref{fig:UAWSoM_T_differentL}. This panel shows how the temperature and velocity at the loop top evolve for this specific case. Slow waves, which seem to be triggered by the instability, are usually damped within several hours. This is consistent with the analysis of \citet{Kolotkov2020} based on the fitted scaling relationship of the heating rate with density and temperature. More specifically, we find that the heating rate per unit mass in the more steady part of UAWSoM simulations follows this relationship very well:
\begin{equation}
    \frac{E_\textrm{H}}{\rho} \propto \rho^a\ T^b
\end{equation}
with the scaling exponents being close to $a\approx -0.85$ and $b\approx -0.15$, with only small deviations between different simulations (up to 0.03 from the values mentioned for both exponents). This also closely matches the profiles obtained by the RTVS-based method (with the only difference that $b\approx -0.10$ in this case).
In this regime, \citet{Kolotkov2020} predict the thermal modes to damp more weakly or be potentially unstable and the acoustic (slow) modes to damp more strongly, which seems consistent with the UAWSoM simulations shown here.

Time periods during which slow waves are damped often coincide with the periods of more pronounced heating and cooling cycles, which is also seen in the left panel of Fig.~\ref{fig:UAWSoM_Tv}. We also noticed that the presence of slow waves in \mbox{UAWSoM} simulations helps in keeping the loops in overall long-term thermal balance, since the periods of undisturbed pronounced heating can lead to `catastrophic heating' episodes described later.

Large-scale oscillations with periods of $0.5-3$ hours are also commonly present. In shorter loops this is typically the dominant feature, clearly seen in the top panel of Fig.~\ref{fig:UAWSoM_T_differentL}, while in longer loops this effect is mixed with the propagation of slow waves, but is also noticeable (middle panel in Fig.~\ref{fig:UAWSoM_T_differentL} and both panels in Fig.~\ref{fig:UAWSoM_T_differentres}). We note that even in shorter loops, slow waves are excited (in the case of the top panel of Fig.~\ref{fig:UAWSoM_T_differentL} after 20000 s); however, the overall loop dynamics is not significantly affected.

These large-scale oscillations show behaviour akin to TNE cycles. The clearest case is seen for the loop which is 50~Mm long (top panel of Fig.~\ref{fig:UAWSoM_T_differentL}). The time evolution of temperature, density and pressure at the loop apex is also shown in Fig.~\ref{fig:UAWSoM_Tprho_16k}. In this case, we find that the peaks in temperature at the loop apex are followed by peaks in kink wave energy density, pressure and finally density with a delay of around $20\%$ of the cycle period, similar to \citet{Mikic2013} and \citet{Froment2018} for a symmetric loop with low heating scaling height. All the aforementioned parameters then reach their minima together with the temperature at the same time, before the loop begins heating. For longer loops where slow waves have a larger influence on the loop dynamics, the overall picture is more complex. This is illustrated in the right panel of Fig.~\ref{fig:UAWSoM_Tv}, which shows the same physical quantities as Fig.~\ref{fig:UAWSoM_Tprho_16k}, but now for the loop of length $2L=150$~Mm. The large-scale oscillations that occur after the first $~7$ hours of the simulation are clearly distinguishable from the short slow-wave-induced oscillations and retain similar characteristics as in the shorter 50-Mm loop. We note that the exact oscillation periods and characteristics can be significantly influenced by the exact numerical setup \citep{Johnston2019}, which is also clear in Fig.~\ref{fig:UAWSoM_T_differentres}. The oscillations in temperature are generally lower than reported in studies on TNE cycles \citep{Auchere2014, Froment2018} and show behaviour more similar to the `incomplete' condensations \citep{Mikic2013}, since in our case the temperature near the apex rarely goes below 0.4~MK. This may be in part due to the heating being inversely proportional to the square root of density, which facilitates the rebound of the loop by increasing the heating per unit mass. However, the choice of the boundary conditions and the lack of chromospheric response likely play a significant role, possibly having a large effect on the development and characteristics of these limit cycles.

One of the main limitations of the current \mbox{UAWSoM} setup is the lack of ability to model very hot loops with temperatures exceeding 2~MK consistently. This is because currently the temperature at the footpoints is allowed to freely adjust in response to the loop conditions. At high enough temperatures, the loop can easily reach a state in which the input heating overtakes the radiative cooling in the whole loop. This inevitably marks the start of the `catastrophic heating' stage, as the excess energy has no place in the domain where it can be distributed. The effect is exacerbated by the fact that the radiative loss function in the solar corona decreases up to around 6~MK, which, together with large mass influx through the footpoints in an attempt to restore the energy balance, leads to a complete breakdown of the initial loop setup. This effect emphasises the importance of the connection between the coronal and chromospheric plasma in a coronal loop and seems to be a direct physical outcome of this connection being broken or severely limited. Overall, keeping the footpoint temperature at a fixed chromospheric value, similar to both the analytical and the RTVS-based method, seems to be essential to extend the range of loops which could be modelled using \mbox{UAWSoM}.

\subsection{Validity of the locality assumption}\label{sect:res_locality}

Before we continue with the comparison of the results obtained by different methods, we wish to answer two unresolved questions regarding the analytical model: how well is the locality assumption $\bar{W}_\textrm{k} \propto T^{-x}$ satisfied, and if so, what value of $x$ should we use in the analytical model. We use the results of the two numerical models for this purpose.

\begin{figure}[ht]
    \centering
    \includegraphics[width=0.9\hsize]{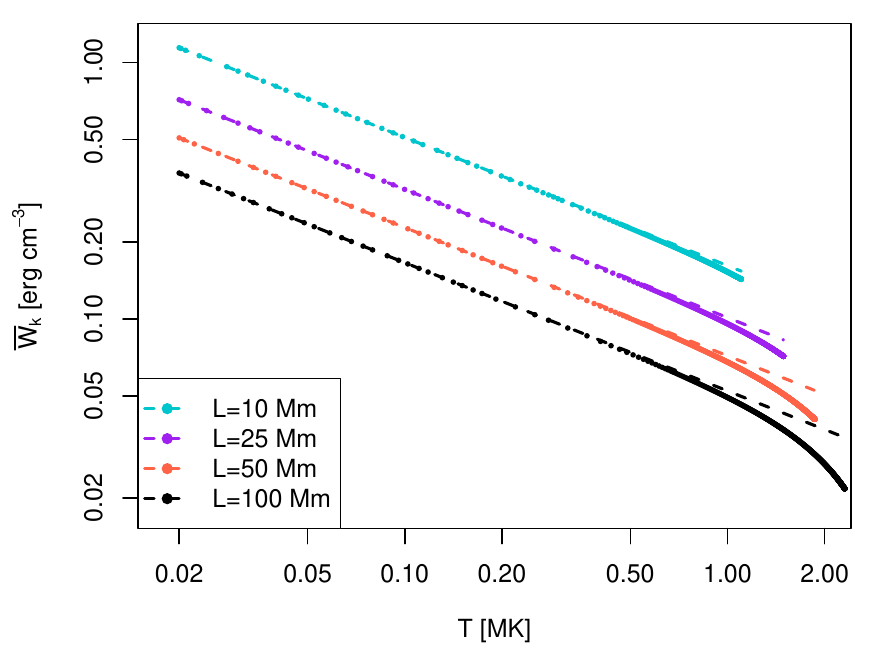}
    \caption{$\bar{W}_\textrm{k}$ vs. $T$ profiles obtained by the RTVS-based method for $P_0=0.5$ dyn cm$^{-2}$ and different loop lengths (denoted by dots) compared to lines with a slope of -0.5 passing through the loop footpoint (denoted by dashed lines). Loops with different loop lengths are denoted by different colours.}
    \label{fig:RTV_fitting}
\end{figure}

In Fig.~\ref{fig:RTV_fitting}, the total $\bar{W}_\textrm{k}$ is plotted as a function of temperature obtained using the RTVS-based approach for a set value of $P_0=0.5$ dyn cm$^{-2}$ and for 4 different loop lengths. In this log-log plot, the deviation from the locality assumption corresponds to the deviation of the data points from a line. Data points in the low-temperature part of the loop (up to around 0.5 MK) follow dashed lines with a slope of -0.5 very closely. This can also be explained by examining the behaviour of the RTVS-based model near the footpoint using Eqs.~\eqref{eq:numeric_Wk} and \eqref{eq:numeric_P}, which is further elaborated in Appendix~\ref{sect:app_numeric_bottom}.

The locality assumption is less accurate for higher temperatures and higher up in the loop, especially as loops become larger. This points to the conclusion that the locality assumption holds the best for smaller loops. We deem that this does not reduce the accuracy and applicability of the analytical results (at least not significantly), as one of the main assumptions in the analytical approach is that the pressure in the loop should be constant, which is only reasonable for loops much smaller than the pressure scale height in the corona anyway.

One can also fit an analytical function to the numerically obtained profiles shown in Fig.~\ref{fig:RTV_fitting} to obtain more insight into the behaviour of the RTVS-based result. The details of these fits are provided in Appendix~\ref{sect:app_Wk_T_fitting}. Overall, we find that parameters $a, b$ and $c$ can be found such that the following expression for $\bar{W}_\textrm{k}$:
\begin{equation}
    \bar{W}_\textrm{k} = T^{-0.5} \exp(a-bT^c)
\end{equation}
approximates the RTVS-based result within 1\% for all loops examined in Fig.~\ref{fig:RTV_fitting}.

\begin{figure}[ht]
    \centering
    \includegraphics[width=0.95\hsize]{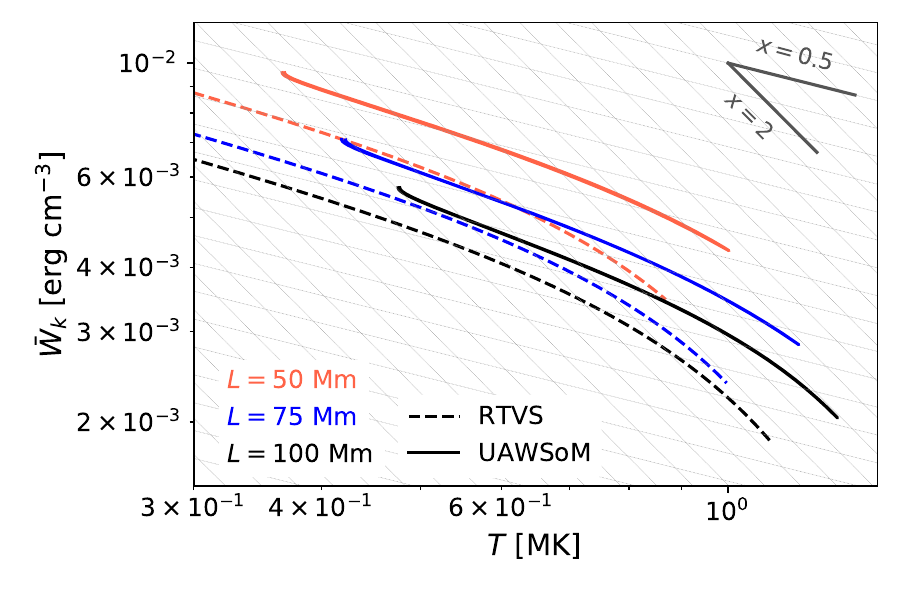}
    \caption{$\bar{W}_\textrm{k}$ vs. $T$ profiles obtained by the RTVS-based method and \mbox{UAWSoM} simulations (denoted by different line types) for loops with different loop lengths (denoted by different colours). UAWSoM results correspond to the time when the loops in the simulations are static and symmetric. Simulations shown here are the same as the ones described and analysed in Sect.~\ref{sect:res_comparison}. The underlying grid shows slopes corresponding to $x=0.5$ and $x=2$ for reference.}
    \label{fig:RTVS_UAWSoM_Wk_T}
\end{figure}

Figure~\ref{fig:RTVS_UAWSoM_Wk_T} shows the same plot, but for individual snapshots from three \mbox{UAWSoM} simulations and their RTVS-based counterparts (meaning that simulations that have the same loop length also have the same pressure at the base), focusing on the upper portion of the loops. The snapshots of the \mbox{UAWSoM} simulations are chosen at a time when the deviations from the loop being symmetric and static are minimal, i.e.\ between the end of the relaxation phase and the first excitation of slow waves. Given that the boundary conditions in \mbox{UAWSoM} are different compared to the RTVS-based model, the limiting behaviour of the \mbox{UAWSoM} results near the footpoints cannot be compared to the RTVS-based model and is, in fact, different. At the same temperature, the value of $x$ in \mbox{UAWSoM} does not seem to show any significant variations and increases from $x=0.7$ near $T=0.4$ MK up to $x=2$ close to $T=1.3$ MK. In RTVS-based results, values of $x$ at the loop top also decrease up to $x=2$. 
Note that $x=2$ is not a limit of any kind in the two numerical approaches; we find that for loops longer than shown in Fig.~\ref{fig:RTVS_UAWSoM_Wk_T}, the value of $x$ can go above 2 near the loop apex. This is in contrast to the analytical method, where the upper limit $x=2$ is strict (recall Eq.~\eqref{eq:integral_condition} and the beginning of Sect.~\ref{sect:res_analytic}).

The discussion of the clear differences between the RTVS-based and \mbox{UAWSoM} results in Fig.~\ref{fig:RTVS_UAWSoM_Wk_T} is deferred to Sect.~\ref{sect:res_comparison}.

\subsection{Comparison between different methods} \label{sect:res_comparison}

We examine the differences between the methods in more detail. Since the analytic and RTVS-based methods only consider a static and symmetric loop, the analysis in this and the following Sect.~\ref{sect:res_scalings} considers only those \mbox{UAWSoM} simulations which go through a phase when the loop is almost completely steady and symmetric (as discussed in Sect.~\ref{sect:res_uawsom} and \ref{sect:res_locality}), and only during this time period. Hence, \mbox{UAWSoM} results here and in Sect.~\ref{sect:res_scalings} represent specific timesteps during which the aforementioned conditions are satisfied. 

\begin{figure}[ht]
    \centering
    \includegraphics[width=0.95\hsize]{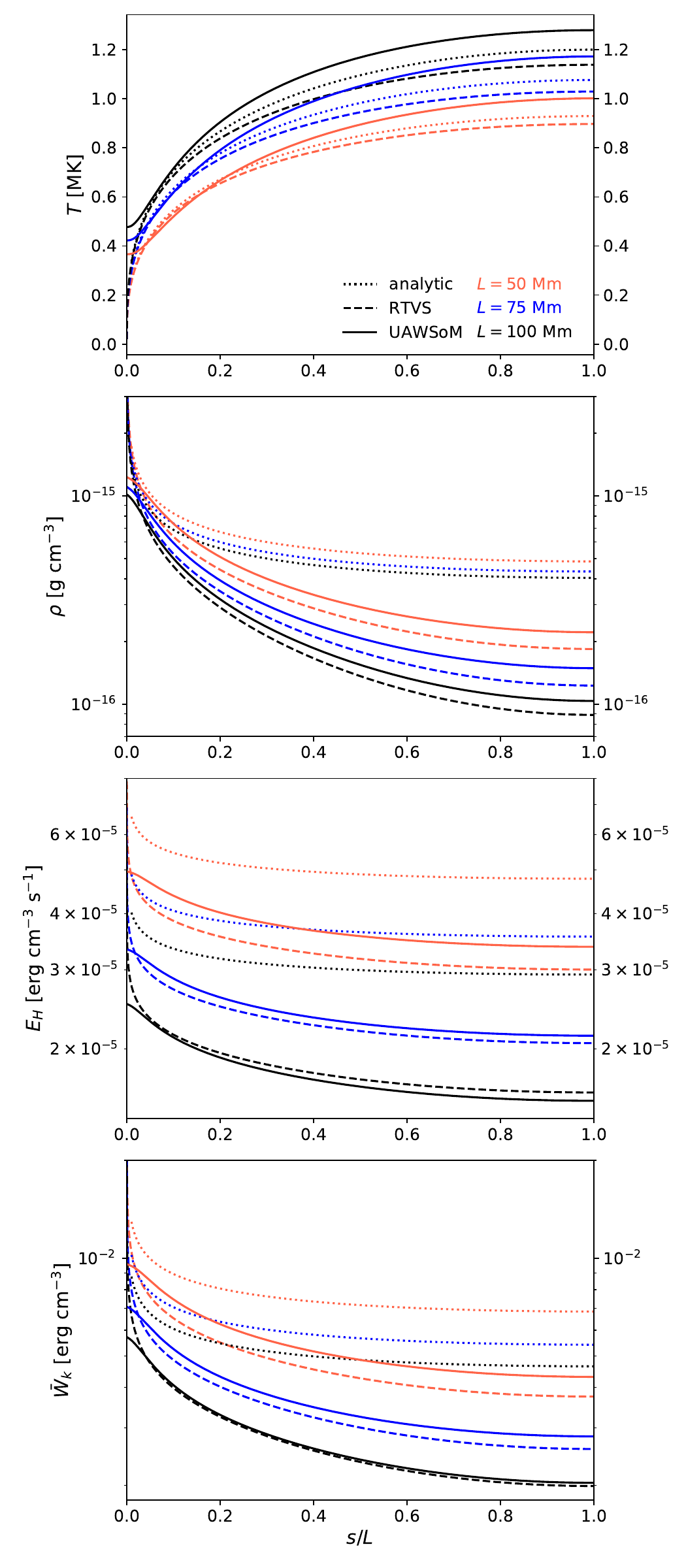}
    \caption{From top to bottom: temperature, density, heating rate and total $\bar{W}_\textrm{k}$ along the loop as predicted by all three methods considered (denoted by different line types) and for different loop lengths (denoted by different colours).}
    \label{fig:comparison_general_all}
\end{figure}

\begin{figure}[ht]
    \centering
    \includegraphics[width=0.95\hsize]{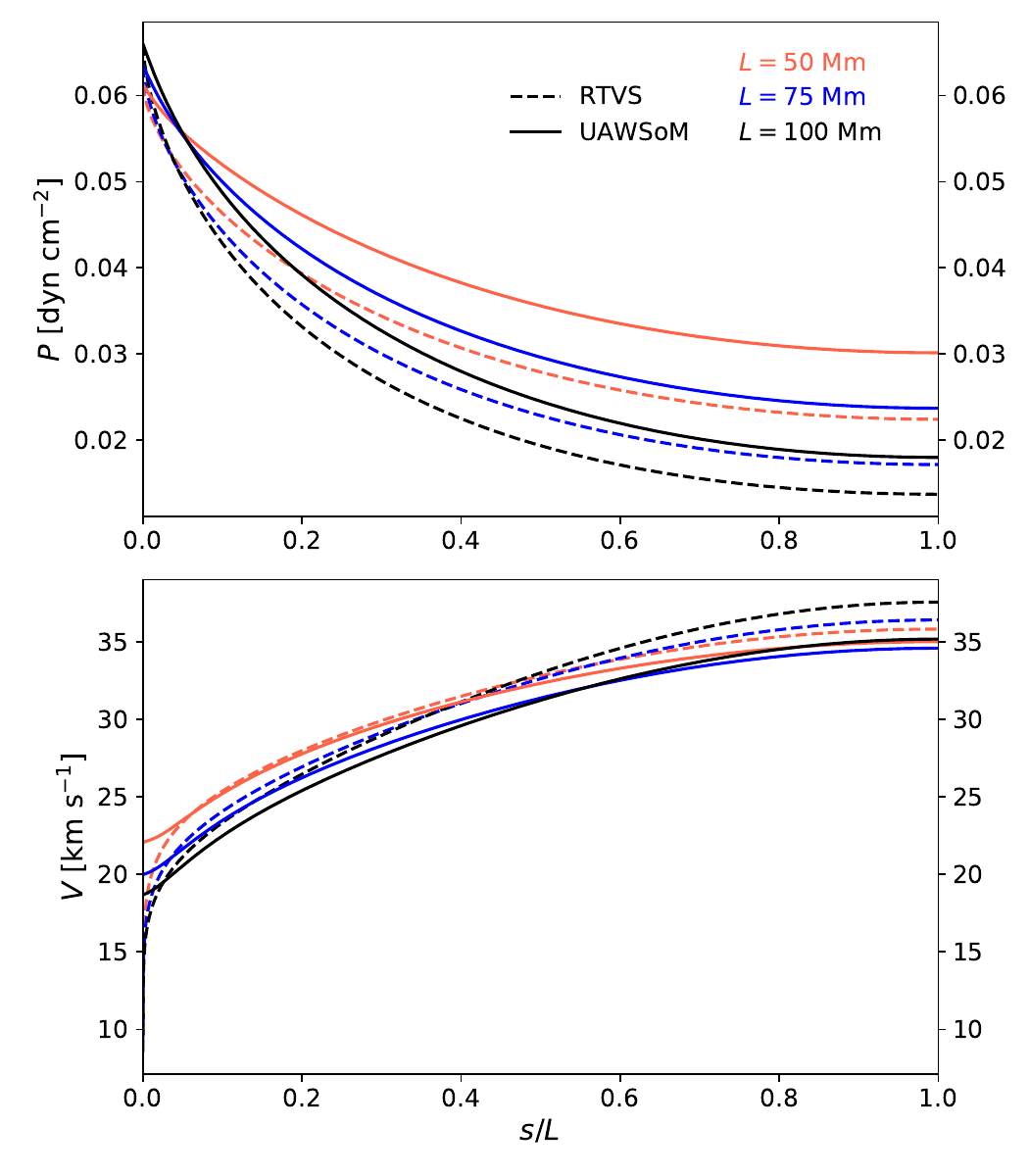}
    \caption{Pressure (top) and velocity amplitude (bottom) along the loop as predicted by the two numerical methods considered (denoted by different line types) and for different loop lengths (denoted by different colours).}
    \label{fig:comparison_general_numeric}
\end{figure}

Figures~\ref{fig:comparison_general_all} and \ref{fig:comparison_general_numeric} show the results of different methods together. Figure~\ref{fig:comparison_general_all} includes temperature, density, heating rate and total $\bar{W}_\textrm{k}$ for all three methods, while Fig.~\ref{fig:comparison_general_numeric} includes pressure and velocity amplitude for the two numerical approaches. The analytical approach is not included in Fig.~\ref{fig:comparison_general_numeric} since pressure is assumed constant there and the velocity amplitude does not come out of the approach straightforwardly since the separate wave energy components $W_\textrm{k}^\pm$ are not modelled. The velocity amplitude is calculated from density and wave energy densities using the following relationship \citep[which follows from e.g.][]{Petrova2023}:
\begin{equation}
    W_\textrm{k}^- + W_\textrm{k}^+ = \frac{1}{2}\left(\rho_e+\rho_i\right)V^2,
\end{equation}
which can be rewritten as
\begin{equation}
    W_\textrm{k}^- + W_\textrm{k}^+ = \frac{1}{2}\frac{\zeta+1}{f\zeta+1-f}\,\rho V^2,
\end{equation}
using the definition of average density.

The simulations whose results are depicted in these two figures are chosen as follows. Regarding the \mbox{UAWSoM} simulations, the loop with half-length $L=50$~Mm has $W_\textrm{k0} = 7\cdot 10^{-3}$ erg cm$^{-3}$, and for the rest of the \mbox{UAWSoM} simulations the value of $W_\textrm{k0}$ is scaled as $L^{-2/3}$, according to Eq.~\eqref{eq:Wk_gen_Martens_Sun}. This is done to keep the base pressure as similar as possible across different simulations. The base pressure $P_0$ in the RTVS-based model and the constant pressure $P$ in the RTV-based derivation are then chosen to match the footpoint pressure of the \mbox{UAWSoM} simulations. In all models we set $A_\textrm{He}=0.1$; note that this differs from the value chosen in Sect.~\ref{sect:res_analytic} and \ref{sect:res_numeric} where no helium was assumed. This change has a small but noticeable impact on the predicted values, which depend on the exact variable and can be up to 15\% for the heating rate.

While the predictions by all three methods are generally similar, we outline and comment on several discrepancies visible in Figs.~\ref{fig:comparison_general_all} and \ref{fig:comparison_general_numeric}. We start with the analytical results, where the assumption that the pressure is constant plays a key role in the results. The typical and well-documented overpressure of the loops using analytical methods \citep[e.g.][]{Aschwanden2001} increases density and radiative losses strongly compared to the other methods, overestimating how much wave energy is needed to heat the loop by around a factor of 2, increasing with loop length. Overall, the agreement with the numerical methods is best for smaller loops; we found that the difference between the analytical and RTVS-based approach becomes reasonably small when $L<20$~Mm.

The two numerical methods agree with each other better. The main difference stems from the fact that the footpoints in the models are at different temperatures: this is fixed at 0.02~MK for the RTVS-based model, while it is variable for \mbox{UAWSoM} and usually settles between 0.2 and 0.6~MK. This impacts the balance between radiative cooling and kink wave heating, as the RTVS-based model additionally includes the part of the domain where cooling dominates. The result is that the pressure and density gradients are larger in the RTVS-based model, which changes the radiative cooling profile significantly. The heating rates and wave energy densities remain similar nonetheless. The total difference between radiative cooling and kink wave heating in the whole domain is always equal to zero.

The velocity amplitudes predicted by the numerical methods are remarkably similar and depend only weakly on loop length (and significantly more on base pressure). For the loops considered in Fig.~\ref{fig:comparison_general_numeric}, the velocity amplitude near the loop apex is predicted to be in the range of 34-40~km s$^{-1}$ and slightly below 10~km s$^{-1}$ near the chromospheric footpoint. We found that for the RTVS-based method $V$ can go up to 50 km s$^{-1}$ for loops with $P_0$ up to 1 dyn cm$^{-2}$. These values are significantly higher than typically observed for decayless transverse oscillations \citep[0.5-20~km s$^{-1}$;][]{Anfinogentov2015, Gao2022, Shrivastav2024, Shrivastav2025}, though values 2-3 times higher than our predictions have also been reported \citep{Petrova2023}. In the case of decaying oscillations, our results fall well within the range of reported amplitudes \citep[][]{Nechaeva2019}. We note, however, that the kink waves which would be the major contributors to the heating would fall within the high-frequency regime and thus may not be directly observable.

This difference in how the balance between heating and cooling is achieved by the numerical methods has a large impact on the predicted maximum temperature, with the \mbox{UAWSoM} typically predicting 10\%-20\% higher loop-top temperature. This is somewhat surprising as the analytical result does not deviate that significantly from the RTVS-based one, and also because temperature is one of the most robust plasma parameters in similar studies considering different heating mechanisms \citep[such as][]{Martens2010}. The temperature discrepancy is consistent with the pressure and density differences due to the difference in energy balance. However, that alone does not seem to provide the complete explanation. Our tests in which we increased the footpoint temperature of the RTVS-based model, which we expected to help bridge this gap, actually resulted in lower maximum temperatures. 

Additionally, we found that we are unable to produce stable loop solutions using the RTVS-based model when we increase the footpoint temperatures to 0.3~MK or more while using the same footpoint pressures as in \mbox{UAWSoM}. This could be viewed as an agreement with the apparent lack of static loop solutions when using \mbox{UAWSoM}. The finding also suggests that the inclusion of the chromosphere in \mbox{UAWSoM} might have a stabilising effect on the loop evolution \citep[which is also supported by the recent study of][]{McMurdo2026b} and thus reduce the parameter space in which slow waves are excited and limit cycles are present. However, we point out that the assumptions of the RTVS-based model, namely the diminishing thermal conduction flux at the footpoints, break down in this case of significantly increased footpoint temperature. Hence, it remains unclear whether the inclusion of the chromosphere in the 1D setting of \mbox{UAWSoM} would indeed eliminate some or all of the dynamic loop behaviour predicted by the \mbox{UAWSoM} simulations.

\subsection{Scaling relationships} \label{sect:res_scalings}

\begin{figure}[ht]
    \centering
    \subfloat[RTVS]{\includegraphics[width=0.49\hsize]{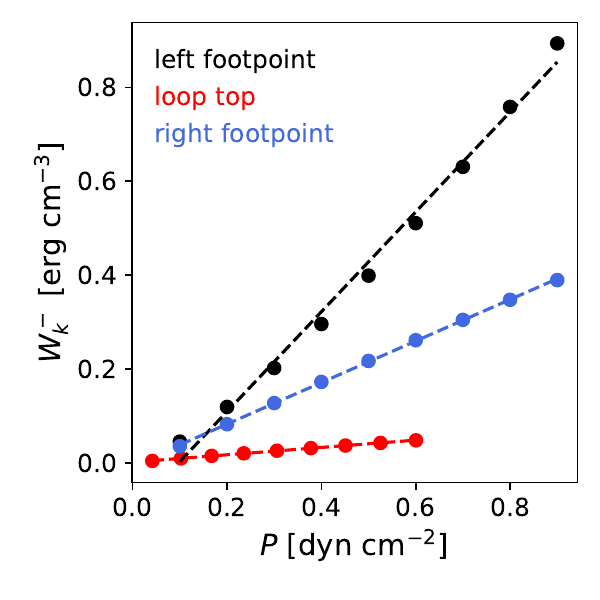}}    
    \subfloat[UAWSoM]{\includegraphics[width=0.49\hsize]{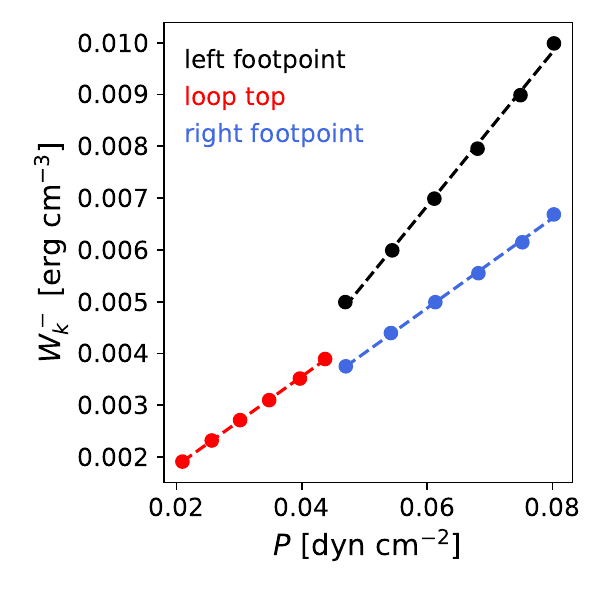}}

    \caption{$W_\textrm{k}^-$ vs. pressure for results obtained using RTVS-based method (left) and \mbox{UAWSoM} (right). All the simulations are of loops with half-length $L=50$ Mm. Linear fits to the data points for each respective location inside the loop (represented by different colours) are shown with dashed lines.}
    \label{fig:scaling_Wk_P_linear} 
\end{figure}

In this part, we investigate how the scaling relationships between kink wave energy density on one side and pressure and loop length on the other change depending on the method used.
Firstly, we test the analytical prediction that the relationship $\bar{W}_\textrm{k} \propto P$ suggested by Eq.~\eqref{eq:Wk_gen_Martens_Sun}. Figure~\ref{fig:scaling_Wk_P_linear} shows the relationship specifically between a single kink wave energy density component $W_\textrm{k}^-$ (corresponding to wave energy propagating from the "left" towards the "right" footpoint) and local pressure, in different parts of the loop and for different numerical methods. All simulated loops have the same length of $2L=100$~Mm. The range of base pressures $P_0$ between 0.04 and 0.1 dyn cm$^{-2}$ is covered by \mbox{UAWSoM}, since this is the range for which stable \mbox{UAWSoM} solutions can be obtained, while the RTVS-based method covers a wider range of base pressures between 0.1 and 0.9 dyn cm$^{-2}$ with increments of 0.1 dyn cm$^{-2}$. For the \mbox{UAWSoM} simulations, the input parameter $W_\textrm{k0}$ is set to the range of 0.05 to 0.1 erg cm$^{-3}$ with increments of 0.01 erg cm$^{-3}$. This setting results in the aforementioned range of footpoint pressure values. Similarly to Sect.~\ref{sect:res_comparison}, only specific timesteps of the UAWSoM simulations are shown here. Note that the large difference in $W_\textrm{k}^-$ values at the footpoints in the RTVS and UAWSoM models stems primarily from the footpoints having significantly different temperatures.

Regardless of the method used, the linear relationship between the quantities seems to be very well satisfied near the loop top and right footpoint. Different slopes in the figure can be explained by the fact that local values of $x$ and values of $T(s)/T_\textrm{max}$ vary along the loop and across different models. Nevertheless, at the loop apex both numerical methods agree on the slope:
\begin{equation}
    W_\textrm{k,top}^\pm \approx 0.087\,P_\textrm{top}.
\end{equation}
This is consistent with the analytical prediction in Eq.~\eqref{eq:Wk_gen_Martens_Sun}, since at the loop apex $T(L)/T_\textrm{max} = \tau(L) = 1$, meaning that the slope at the loop apex should not depend on the method used, as long as the value of $x$ remains similar.

The slopes at the footpoint where wave energy exits cannot be meaningfully computed using analytical predictions, since the wave energy value at this point is predominantly governed by the injected energy at the other footpoint and loop length. In any case, we wish to point out that the analytical scaling $\bar{W}_\textrm{k} \propto P$ seems to hold well enough even beyond the scope for which it was predicted, including both individual and total wave energy components and throughout large parts of the loop.

\begin{figure}[ht]
    \centering
    \subfloat[RTVS]{\includegraphics[width=0.8\hsize]{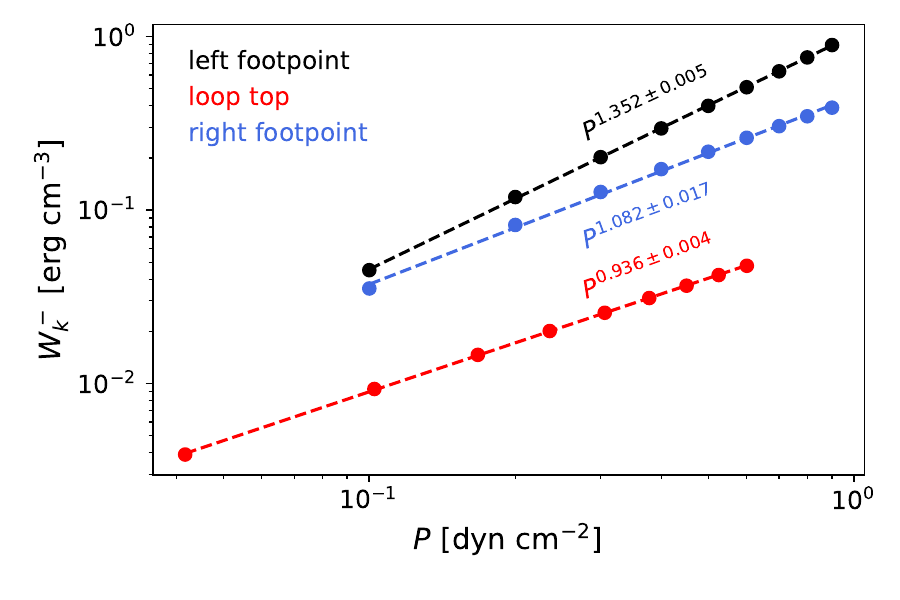}}
    
    \subfloat[UAWSoM]{\includegraphics[width=0.8\hsize]{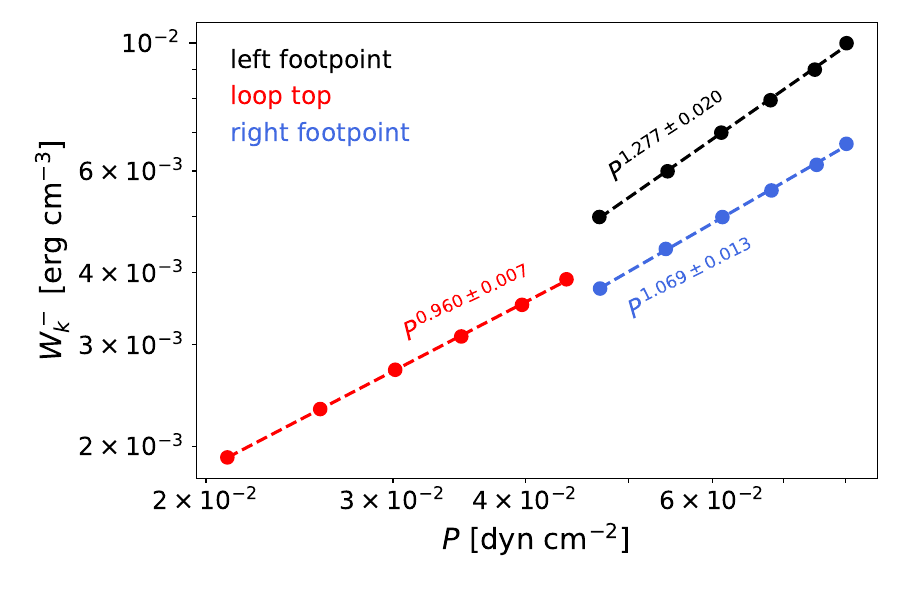}}

    \caption{Similar to Fig.~\ref{fig:scaling_Wk_P_linear}, but with the data now being shown in the log-log plot and the linear fits applied after the axes transformation to estimate the scaling exponent. The resulting scaling fits are represented by dashed lines. The scaling exponents and their standard errors are also denoted for each of the fits.}
    \label{fig:scaling_Wk_P} 
\end{figure}

The quality of the analytical prediction is tested more rigorously by estimating the scaling exponent in the relationship between $W_\textrm{k}^-$ and $P$, the results of which are shown in Fig.~\ref{fig:scaling_Wk_P}. The scaling exponents at the loop apex and right footpoint are indeed very close to unity and almost equal between the two numerical methods. The scaling exponent attains higher values (between 1.25 and 1.36) near the left footpoint, where the wave energy is entering the loop.

Similar scaling relationships also hold for total $\bar{W}_\textrm{k}$ as defined in Eq.~\eqref{eq:Wk_tot_definition}. The scaling laws near the footpoint can be summarised as follows:
\begin{equation}\label{eq:scaling_add1}
    \bar{W}_\textrm{k}\propto \begin{cases}
        P^{1.26}, & \text{RTVS-based model},\\
        P^{1.20}, & \text{UAWSoM},\\
    \end{cases}
\end{equation}
while near the loop apex the scaling exponents are equal to the ones shown in Fig.~\ref{fig:scaling_Wk_P} (which is a direct consequence of the definition of $\bar{W}_\textrm{k}$ in terms of wave energy components and loop symmetry). Since the results are obtained using the same method and are similar to the $W_\textrm{k}^-$ case, the corresponding plots showing the fits can be consulted in Appendix~\ref{sect:app_scaling}, Fig.~\ref{fig:scaling_app_Wktot_P}.

\begin{figure}[ht]
    \centering
    \subfloat[RTVS]{\includegraphics[width=0.8\hsize]{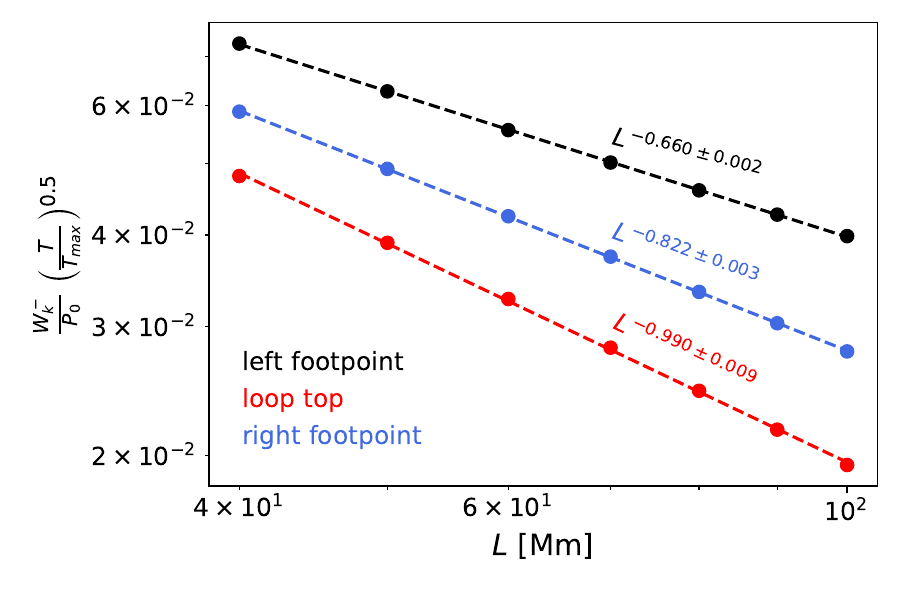}}
    
    \subfloat[UAWSoM]{\includegraphics[width=0.8\hsize]{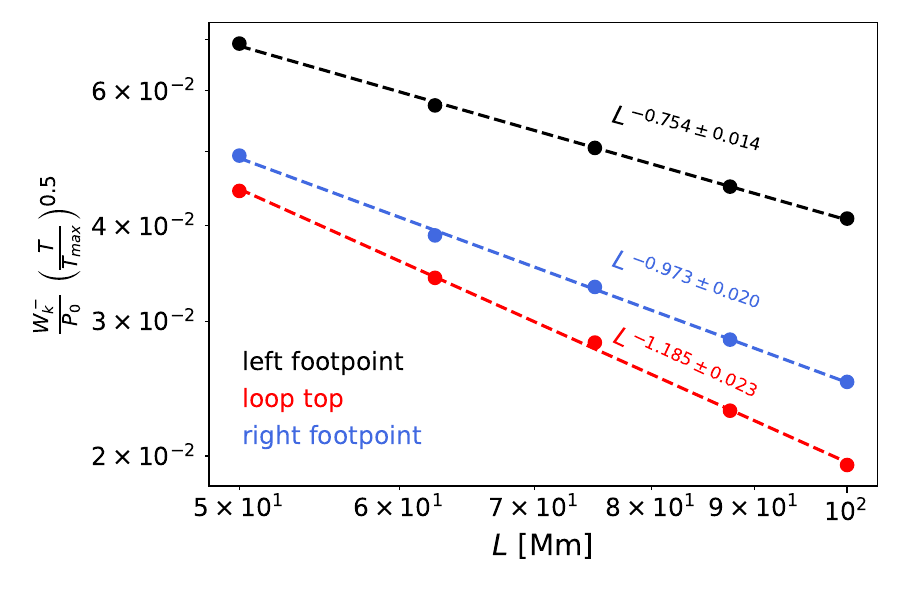}}

    \caption{Scaling laws between appropriately scaled $W_\textrm{k}^-$ and $L$ for the results obtained using the RTVS-based method (top) and \mbox{UAWSoM} (bottom). Fitted scaling laws are represented by dashed lines, each corresponding to the specific part of the loop.}
    \label{fig:scaling_Wk_L} 
\end{figure}

Secondly, we investigate the relationship between $\bar{W}_\textrm{k}$ and $L$. To obtain a scaling relationship, it is necessary to appropriately rescale $\bar{W}_\textrm{k}$ so that all other important contributions in Eq.~\eqref{eq:Wk_gen_Martens_Sun} are correctly accounted for. This includes the factor $\tau(s)^{-x} = \left(T(s)/T_\textrm{max}\right)^{-x}$, which is the only factor in the scaling law that changes along the loop, and pressure $P$, which remains to be a constant as to not introduce another factor that changes along the loop, and is chosen to be equal to the footpoint pressure $P_0$ since it provides the best results.

Hence, Fig.~\ref{fig:scaling_Wk_L} shows the relationship between the scaled single wave energy density component, which is the dimensionless quantity $\frac{W_\textrm{k}^-}{P_0} \left(\frac{T}{T_\textrm{max}}\right)^{1/2}$, and half-length $L$. Note that we choose $x=0.5$ here based on the discussion in Sect.~\ref{sect:res_locality} on the behaviour of the RTVS-based model close to the footpoints, and the fact that this choice does not affect the loop top, since there $T = T_\textrm{max}$. Simulation data and scaling fits for both numerical methods and for both footpoints and at the loop top are shown. Again, for \mbox{UAWSoM} the results shown are chosen from the time period when the loop is approximately steady and symmetric. In the RTVS-based results, the base pressure is set to $P_0 = 0.1$ dyn cm$^{-2}$ in all cases. In UAWSoM simulations, we set $W_\textrm{k0}=0.07$ erg cm$^{-3}$ for the loop of length $2L=100$ Mm; for loops of other lengths this value is adjusted to scale with $L^{-2/3}$ as with the simulations in Sect.~\ref{sect:res_comparison}. In this manner we try to keep footpoint pressure across different UAWSoM simulations as similar as possible.

The scaling fits are compared to the equivalent analytical prediction (obtained by rewriting Eq.~\eqref{eq:Wk_gen_Martens_Sun}), which suggests that $\frac{\bar{W}_\textrm{k}}{P} \left(\frac{T}{T_\textrm{max}}\right)^{x} \propto L^{-2/3}$ holds. Note that in the numerical models we test a slightly different quantity $\frac{W_\textrm{k}^-}{P_0} \left(\frac{T}{T_\textrm{max}}\right)^{1/2}$, where we substitute $P$ specifically by $P_0$, $x$ by $0.5$ and $\bar{W}_\textrm{k}$ by just a single propagating component $W_\textrm{k}^-$. By specifying $P$ and $x$ in this manner, we expect to match the exponent obtained by the scaling fit to the RTVS-based model near the left footpoint where the wave energy component $W_\textrm{k}^-$ is entering the domain with the analytically predicted exponent of $-2/3$. This is indeed the case, which is encouraging. Elsewhere in the loop and in \mbox{UAWSoM}, however, the scaling exponent is lower than -2/3, with the values as low as -1.2 at the loop top in \mbox{UAWSoM}. The scaling exponents are denoted exactly in Fig.~\ref{fig:scaling_Wk_L}. Again, the agreement between the analytical and numerical models seems to be the best in the low-temperature regime near the footpoints.

We wish to add two additional remarks. In the case of the RTVS-based model, we noticed that near both footpoints, $T/T_\textrm{max}$ scales roughly as $L^{-1/3}$. The consequence of this is that the scaling laws
\begin{equation}\label{eq:scaling_add2}
    \frac{W_\textrm{k}^-}{P_0} \propto \begin{cases}
        L^{-1/2}, & \text{near the left footpoint},\\
        L^{-1}, & \text{near the loop top},\\
        L^{-2/3}, & \text{near the right footpoint},
    \end{cases}
\end{equation}
represent the RTVS-based results extremely well; the scaling exponents obtained by fitting the data with the scaling laws differ from the exponents given above by at most 1\%. In the case of \mbox{UAWSoM}, the ratio $T/T_\textrm{max}$ changes only a little between different simulations, meaning that $\frac{W_\textrm{k}^-}{P_0}$ and $\frac{W_\textrm{k}^-}{P_0} \left(\frac{T}{T_\textrm{max}}\right)^{1/2}$ scale with $L$ with the same scaling exponents already listed in the bottom panel of Fig.~\ref{fig:scaling_Wk_L}. The data and resulting fits are shown in Appendix~\ref{sect:app_scaling}, Fig.~\ref{fig:scaling_app1}.

The other remark is in regard to the scaling of the total $\bar{W}_\textrm{k}$ with $L$. The scaling exponent does not change compared to $W_\textrm{k}^-$ (or $W_\textrm{k}^+$) at the loop top due to symmetry. However, it decreases near the footpoints compared to the scaling of the entering individual wave energy component ($W_\textrm{k}^-$ at the left and $W_\textrm{k}^+$ at the right footpoint; Fig.~\ref{fig:scaling_Wk_L}) as follows:
\begin{equation}\label{eq:scaling_add3}
    \frac{\bar{W}_\textrm{k}}{P_0} \left(\frac{T}{T_\textrm{max}}\right)^{1/2}\propto \begin{cases}
        L^{-0.72}, & \text{RTVS-based model},\\
        L^{-0.83}, & \text{UAWSoM}.\\
    \end{cases}
\end{equation}
The decrease is similar for $\frac{\bar{W}_\textrm{k}}{P_0}$ when compared to the results in Fig.~\ref{fig:scaling_Wk_P}. The exact fits are shown in Fig.~\ref{fig:scaling_app_Wktot_L} in Appendix~\ref{sect:app_scaling}.

\section{Conclusions} \label{sect:conclusion}

In this work, we performed a 1D analytical and numerical analysis of coronal loops heated primarily by turbulent dissipation of kink waves in the solar corona. We focused on comparing the three different methods, each with its own set of assumptions and computational complexity, to arrive at estimates of heating rates and kink wave energy densities needed to support the coronal loops against radiative cooling.

The analytical method generalises the RTV scaling laws in this setting and is the simplest to use. It can be used to derive reasonable estimates of the major scaling relationships between different coronal loop parameters. These are, namely, $T_\textrm{max} \propto (PL)^{1/3}, E_\textrm{H}(s) \propto  (T(s)/T_\textrm{max})^{(1-3x)/2}\, P^{7/6}\,L^{-5/6}$ and $\bar{W}_\textrm{k}(s) \propto (T(s)/T_\textrm{max})^{-x}\,P\,L^{-2/3}$. The assumption of constant pressure causes the well-documented overestimation of density, in our case by a factor between 2 and 5 at the loop apex, increasing with loop length. The resulting radiative loss increase is counteracted by overestimating the heating rate and kink wave energy density compared to the numerical predictions. The extent of this effect somewhat depends on the chosen value of parameter $x$; however, the discrepancies in the heating profiles between the analytical and the two numerical methods cannot be fully eliminated. As a result, the range in which the method provides accurate results is constrained to short loops with loop lengths below $2L \lesssim 40$~Mm.

The RTVS-based numerical method seems to be the most useful of the three, given its minimal computational cost needed to yield more realistic coronal loop profiles for a wide range of loops. Therefore, we also make the code of this method publicly available and welcome any suggestions regarding its further improvements.

Current results of coronal loop modelling using \mbox{UAWSoM} suggest that the loops may be prone to being thermally unstable and developing TNE cycles, in contrast to the stable regimes predicted by the RTVS-based method. However, more work on improving \mbox{UAWSoM} in this context is needed. Using \mbox{UAWSoM} in the scope of this study helped in recognising some of the remaining challenges in the ongoing development of the 3D version of \mbox{UAWSoM}. In particular, the inclusion of the chromosphere in the simulation domain may be necessary to ensure the long-term stability of the model, despite the associated increase in computational costs.

Overall, when comparing the three methods between themselves, we found that the assumption of constant pressure along the loop severely limits the accuracy and applicability of the analytical approach. The locality assumption, which links the kink wave energy density to local plasma properties, is found to be another strong assumption whose range of validity is similar to the assumption of constant pressure. The usefulness of the analytical approach remains in the prediction of scaling law exponents, which agree reasonably with the numerical results. The two numerical approaches agree with each other well, with the differences in the predicted loop apex quantities limited to 10\% for velocity amplitudes, 15\% for heating rate and kink wave energy density, and 20\% for temperature. The discrepancies between them can be primarily linked to the different numerical setup (in particular with regard to the location of the footpoint(s) and whether the loop is assumed to be static or not). It is possible, however, that the additional wave pressure terms in the UAWSoM equations have a direct impact on the existence of the static 1D loop solutions.

At the same time, we recognise that a 1D study in which only a single novel time-dependent model was used is not enough to give a definitive answer to the question of whether transverse oscillations may be responsible for providing most of the heating to coronal loops. The 2.5D study of \citet{McMurdo2026b} is already an important step forward. Still, more advanced 3D models are needed in which the heating will be modelled in the same way as here or possibly in combinations with different heating mechanisms. On the other hand, a successful link from the analytical and numerical results to observations is crucial. So far it has proven difficult, if not impossible, to pinpoint the primary heating source by observing the temperature and density of coronal loop plasma and tracking its dynamics. Kink wave energy and velocity amplitude estimates given in this paper, together with the power laws provided by the analytical method and derived from two numerical methods in Sect.~\ref{sect:res_scalings}, already provide a wide array of results that can be used to compare with coronal loop observations.

At the end, we want to point out a hidden caveat in the assumed heating term we use throughout this paper, namely that this term has been derived for propagating kink waves. While propagating kink waves have been observed in coronal loops \citep{Morton2021, Tiwari2021}, standing kink waves are also present \citep[e.g. review by][]{Nakariakov2021}. The reason the heating term for propagating waves has been used in this study is due to the fact that the equivalent term for standing waves is yet to be derived. Besides, it is also the term which is currently used in \mbox{UAWSoM}. The main difference compared to the heating term for propagating waves would be that the heating would be more localised \citep{VanDoorsselaere2020}, with unclear effects on the loop dynamics. We also point out that \mbox{AWSoM} has been used to study coronal loops with unchanged heating rates compared to the open field \citep{Shi2024}. Further investigation into the energy contribution from standing kink waves and their effect on coronal loops is necessary.

\begin{acknowledgements}
      LB, TVD and MM received financial support from the Flemish Government under the long-term structural Methusalem funding program, project SOUL: Stellar evolution in full glory, grant METH/24/012 at KU Leuven. H.W. is supported by JSPS KAKENHI grant Nos. 599 JP24KJ0151 and JP26K17205. The research that led to these results was subsidised by the Belgian Federal Science Policy Office through the contract B2/223/P1/CLOSE-UP. It is also part of the DynaSun project and has thus received funding under the Horizon Europe programme of the European Union under grant agreement (no. 101131534). Views and opinions expressed are however those of the author(s) only and do not necessarily reflect those of the European Union and therefore the European Union cannot be held responsible for them.
\end{acknowledgements}

\bibliographystyle{aa.bst}
\bibliography{bib.bib}

\begin{thebibliography}{79}
\expandafter\ifx\csname natexlab\endcsname\relax\def\natexlab#1{#1}\fi

\bibitem[{{Anfinogentov} {et~al.}(2015){Anfinogentov}, {Nakariakov}, \&
  {Nistic{\`o}}}]{Anfinogentov2015}
{Anfinogentov}, S.~A., {Nakariakov}, V.~M., \& {Nistic{\`o}}, G. 2015, \aap,
  583, A136

\bibitem[{{Antolin} \& {Froment}(2022)}]{Antolin2022}
{Antolin}, P. \& {Froment}, C. 2022, Frontiers in Astronomy and Space Sciences,
  9, 820116

\bibitem[{{Aschwanden} {et~al.}(1999){Aschwanden}, {Fletcher}, {Schrijver}, \&
  {Alexander}}]{Aschwanden1999}
{Aschwanden}, M.~J., {Fletcher}, L., {Schrijver}, C.~J., \& {Alexander}, D.
  1999, \apj, 520, 880

\bibitem[{{Aschwanden} {et~al.}(2000){Aschwanden}, {Nightingale}, \&
  {Alexander}}]{Aschwanden2000}
{Aschwanden}, M.~J., {Nightingale}, R.~W., \& {Alexander}, D. 2000, \apj, 541,
  1059

\bibitem[{{Aschwanden} \& {Schrijver}(2002)}]{Aschwanden2002}
{Aschwanden}, M.~J. \& {Schrijver}, C.~J. 2002, \apjs, 142, 269

\bibitem[{{Aschwanden} {et~al.}(2001){Aschwanden}, {Schrijver}, \&
  {Alexander}}]{Aschwanden2001}
{Aschwanden}, M.~J., {Schrijver}, C.~J., \& {Alexander}, D. 2001, \apj, 550,
  1036

\bibitem[{{Asgari-Targhi} \& {van Ballegooijen}(2025)}]{Asgari-Targhi2025}
{Asgari-Targhi}, M. \& {van Ballegooijen}, A.~A. 2025, \apj, 985, 96

\bibitem[{{Auch{\`e}re} {et~al.}(2014){Auch{\`e}re}, {Bocchialini}, {Solomon},
  \& {Tison}}]{Auchere2014}
{Auch{\`e}re}, F., {Bocchialini}, K., {Solomon}, J., \& {Tison}, E. 2014, \aap,
  563, A8

\bibitem[{{Bennedik} {et~al.}(2026){Bennedik}, {Stelzer}, {Caramazza},
  {Orlando}, {Binks}, \& {Robrade}}]{Bennedik2026submitted}
{Bennedik}, M.~M., {Stelzer}, B., {Caramazza}, M., {et~al.} 2026, \aap, in
  press

\bibitem[{{Bourdin} {et~al.}(2016){Bourdin}, {Bingert}, \&
  {Peter}}]{Bourdin2016}
{Bourdin}, P.-A., {Bingert}, S., \& {Peter}, H. 2016, \aap, 589, A86

\bibitem[{{Bradshaw} \& {Emslie}(2020)}]{Bradshaw2020}
{Bradshaw}, S.~J. \& {Emslie}, A.~G. 2020, \apj, 904, 141

\bibitem[{{Bradshaw} {et~al.}(2019){Bradshaw}, {Emslie}, {Bian}, \&
  {Kontar}}]{Bradshaw2019}
{Bradshaw}, S.~J., {Emslie}, A.~G., {Bian}, N.~H., \& {Kontar}, E.~P. 2019,
  \apj, 880, 80

\bibitem[{{Brooks} {et~al.}(2021){Brooks}, {Warren}, \& {Landi}}]{Brooks2021}
{Brooks}, D.~H., {Warren}, H.~P., \& {Landi}, E. 2021, \apjl, 915, L24

\bibitem[{{\v{C}}ada \& Torrilhon(2009)}]{Cada2009}
{\v{C}}ada, M. \& Torrilhon, M. 2009, Journal of Computational Physics, 228,
  4118

\bibitem[{{Cargill} {et~al.}(2016){Cargill}, {De Moortel}, \&
  {Kiddie}}]{Cargill2016}
{Cargill}, P.~J., {De Moortel}, I., \& {Kiddie}, G. 2016, \apj, 823, 31

\bibitem[{{Colgan} {et~al.}(2008){Colgan}, {Abdallah}, {Sherrill}, {Foster},
  {Fontes}, \& {Feldman}}]{Colgan2008}
{Colgan}, J., {Abdallah}, Jr., J., {Sherrill}, M.~E., {et~al.} 2008, \apj, 689,
  585

\bibitem[{{Cranmer} \& {Winebarger}(2019)}]{Cranmer2019}
{Cranmer}, S.~R. \& {Winebarger}, A.~R. 2019, \araa, 57, 157

\bibitem[{{Dai} {et~al.}(2024){Dai}, {Xiang}, \& {Ding}}]{Dai2024}
{Dai}, Y., {Xiang}, J.~J., \& {Ding}, M.~D. 2024, \apj, 965, 2

\bibitem[{{Downs} {et~al.}(2016){Downs}, {Lionello}, {Miki{\'c}}, {Linker}, \&
  {Velli}}]{Downs2016}
{Downs}, C., {Lionello}, R., {Miki{\'c}}, Z., {Linker}, J.~A., \& {Velli}, M.
  2016, \apj, 832, 180

\bibitem[{{Elsässer}(1950)}]{Elsasser1950}
{Elsässer}, W.~M. 1950, Physical Review, 79, 183

\bibitem[{{Emslie} \& {Bradshaw}(2022)}]{Emslie2022}
{Emslie}, A.~G. \& {Bradshaw}, S.~J. 2022, \apj, 939, 19

\bibitem[{{Froment} {et~al.}(2018){Froment}, {Auch{\`e}re}, {Miki{\'c}},
  {Aulanier}, {Bocchialini}, {Buchlin}, {Solomon}, \&
  {Soubri{\'e}}}]{Froment2018}
{Froment}, C., {Auch{\`e}re}, F., {Miki{\'c}}, Z., {et~al.} 2018, \apj, 855, 52

\bibitem[{{Gao} {et~al.}(2022){Gao}, {Tian}, {Van Doorsselaere}, \&
  {Chen}}]{Gao2022}
{Gao}, Y., {Tian}, H., {Van Doorsselaere}, T., \& {Chen}, Y. 2022, \apj, 930,
  55

\bibitem[{{Hermans} \& {Keppens}(2021)}]{Hermans2021}
{Hermans}, J. \& {Keppens}, R. 2021, \aap, 655, A36

\bibitem[{{Heyvaerts} \& {Priest}(1983)}]{Heyvaerts1983}
{Heyvaerts}, J. \& {Priest}, E.~R. 1983, \aap, 117, 220

\bibitem[{{Hollweg}(1986)}]{Hollweg1986}
{Hollweg}, J.~V. 1986, \jgr, 91, 4111

\bibitem[{{Hood} \& {Priest}(1979)}]{Hood1979}
{Hood}, A.~W. \& {Priest}, E.~R. 1979, A\&A, 77, 233

\bibitem[{{Howson} {et~al.}(2020){Howson}, {De Moortel}, \&
  {Reid}}]{Howson2020}
{Howson}, T.~A., {De Moortel}, I., \& {Reid}, J. 2020, \aap, 636, A40

\bibitem[{{Johnston} {et~al.}(2019){Johnston}, {Cargill}, {Antolin}, {Hood},
  {De Moortel}, \& {Bradshaw}}]{Johnston2019}
{Johnston}, C.~D., {Cargill}, P.~J., {Antolin}, P., {et~al.} 2019, \aap, 625,
  A149

\bibitem[{{Kano} \& {Tsuneta}(1995)}]{Kano1995}
{Kano}, R. \& {Tsuneta}, S. 1995, \apj, 454, 934

\bibitem[{{Karampelas} {et~al.}(2017){Karampelas}, {Van Doorsselaere}, \&
  {Antolin}}]{Karampelas2017}
{Karampelas}, K., {Van Doorsselaere}, T., \& {Antolin}, P. 2017, \aap, 604,
  A130

\bibitem[{{Karampelas} {et~al.}(2024){Karampelas}, {Van Doorsselaere}, {Guo},
  {Duckenfield}, \& {Pelouze}}]{Karampelas2024}
{Karampelas}, K., {Van Doorsselaere}, T., {Guo}, M., {Duckenfield}, T., \&
  {Pelouze}, G. 2024, \aap, 688, A80

\bibitem[{{Keppens} {et~al.}(2003){Keppens}, {Nool}, {T{\'o}th}, \&
  {Goedbloed}}]{Keppens2003}
{Keppens}, R., {Nool}, M., {T{\'o}th}, G., \& {Goedbloed}, J.~P. 2003, Computer
  Physics Communications, 153, 317

\bibitem[{{Keppens} {et~al.}(2023){Keppens}, {Popescu Braileanu}, {Zhou},
  {Ruan}, {Xia}, {Guo}, {Claes}, \& {Bacchini}}]{Keppens2023}
{Keppens}, R., {Popescu Braileanu}, B., {Zhou}, Y., {et~al.} 2023, \aap, 673,
  A66

\bibitem[{{Klimchuk} \& {Luna}(2019)}]{Klimchuk2019}
{Klimchuk}, J.~A. \& {Luna}, M. 2019, \apj, 884, 68

\bibitem[{{Kolotkov} {et~al.}(2020){Kolotkov}, {Duckenfield}, \&
  {Nakariakov}}]{Kolotkov2020}
{Kolotkov}, D.~Y., {Duckenfield}, T.~J., \& {Nakariakov}, V.~M. 2020, \aap,
  644, A33

\bibitem[{Krupka {et~al.}(2026)Krupka, Van~Hoof, \&
  Van~Doorsselaere}]{Krupka2026}
Krupka, A., Van~Hoof, T., \& Van~Doorsselaere, T. 2026, Physics of Plasmas, 33,
  082104

\bibitem[{{Lim} {et~al.}(2024){Lim}, {Van Doorsselaere}, {Berghmans}, \&
  {Petrova}}]{Lim2024}
{Lim}, D., {Van Doorsselaere}, T., {Berghmans}, D., \& {Petrova}, E. 2024,
  \aap, 689, A16

\bibitem[{{Mac Cormack} {et~al.}(2020){Mac Cormack}, {L{\'o}pez Fuentes},
  {Mandrini}, {Lloveras}, {Poisson}, \& {V{\'a}squez}}]{MacCormack2020}
{Mac Cormack}, C., {L{\'o}pez Fuentes}, M., {Mandrini}, C.~H., {et~al.} 2020,
  Advances in Space Research, 65, 1616

\bibitem[{{Magyar} {et~al.}(2017){Magyar}, {Van Doorsselaere}, \&
  {Goossens}}]{Magyar2017}
{Magyar}, N., {Van Doorsselaere}, T., \& {Goossens}, M. 2017, Scientific
  Reports, 7, 14820

\bibitem[{{Magyar} {et~al.}(2019){Magyar}, {Van Doorsselaere}, \&
  {Goossens}}]{Magyar2019}
{Magyar}, N., {Van Doorsselaere}, T., \& {Goossens}, M. 2019, \apj, 882, 50

\bibitem[{{Martens}(2010)}]{Martens2010}
{Martens}, P.~C.~H. 2010, ApJ, 714, 1290

\bibitem[{{McMurdo} {et~al.}(2025){McMurdo}, {Ballai}, {Verth}, \&
  {Fedun}}]{McMurdo2025}
{McMurdo}, M., {Ballai}, I., {Verth}, G., \& {Fedun}, V. 2025, \apj, 988, 50

\bibitem[{{McMurdo} {et~al.}(2026{\natexlab{a}}){McMurdo}, {Jer{\v{c}}i{\'c}},
  {Van Doorsselaere}, \& {Froment}}]{McMurdo2026b}
{McMurdo}, M., {Jer{\v{c}}i{\'c}}, V., {Van Doorsselaere}, T., \& {Froment}, C.
  2026{\natexlab{a}}, \aap, 714, A450

\bibitem[{{McMurdo} {et~al.}(2026{\natexlab{b}}){McMurdo}, {Van Doorsselaere},
  {Magyar}, {Banovi{\'c}}, \& {Lim}}]{McMurdo2026}
{McMurdo}, M., {Van Doorsselaere}, T., {Magyar}, N., {Banovi{\'c}}, L., \&
  {Lim}, D. 2026{\natexlab{b}}, \aap, 705, A15

\bibitem[{{Miki{\'c}} {et~al.}(2013){Miki{\'c}}, {Lionello}, {Mok}, {Linker},
  \& {Winebarger}}]{Mikic2013}
{Miki{\'c}}, Z., {Lionello}, R., {Mok}, Y., {Linker}, J.~A., \& {Winebarger},
  A.~R. 2013, \apj, 773, 94

\bibitem[{{Morton} {et~al.}(2026){Morton}, {Gao}, {Tajfirouze}, {Tian}, {Van
  Doorsselaere}, \& {Schad}}]{Morton2026}
{Morton}, R.~J., {Gao}, Y., {Tajfirouze}, E., {et~al.} 2026, Nature Astronomy,
  10, 42

\bibitem[{{Morton} {et~al.}(2021){Morton}, {Tiwari}, {Van Doorsselaere}, \&
  {McLaughlin}}]{Morton2021}
{Morton}, R.~J., {Tiwari}, A.~K., {Van Doorsselaere}, T., \& {McLaughlin},
  J.~A. 2021, \apj, 923, 225

\bibitem[{{Nakariakov} {et~al.}(2021){Nakariakov}, {Anfinogentov}, {Antolin},
  {Jain}, {Kolotkov}, {Kupriyanova}, {Li}, {Magyar}, {Nistic{\`o}}, {Pascoe},
  {Srivastava}, {Terradas}, {Vasheghani Farahani}, {Verth}, {Yuan}, \&
  {Zimovets}}]{Nakariakov2021}
{Nakariakov}, V.~M., {Anfinogentov}, S.~A., {Antolin}, P., {et~al.} 2021, \ssr,
  217, 73

\bibitem[{{Nakariakov} {et~al.}(1999){Nakariakov}, {Ofman}, {Deluca},
  {Roberts}, \& {Davila}}]{Nakariakov1999}
{Nakariakov}, V.~M., {Ofman}, L., {Deluca}, E.~E., {Roberts}, B., \& {Davila},
  J.~M. 1999, Science, 285, 862

\bibitem[{{Nechaeva} {et~al.}(2019){Nechaeva}, {Zimovets}, {Nakariakov}, \&
  {Goddard}}]{Nechaeva2019}
{Nechaeva}, A., {Zimovets}, I.~V., {Nakariakov}, V.~M., \& {Goddard}, C.~R.
  2019, \apjs, 241, 31

\bibitem[{{Pagano} \& {De Moortel}(2017)}]{Pagano2017}
{Pagano}, P. \& {De Moortel}, I. 2017, \aap, 601, A107

\bibitem[{{Pelouze} {et~al.}(2022){Pelouze}, {Auch{\`e}re}, {Bocchialini},
  {Froment}, {Miki{\'c}}, {Soubri{\'e}}, \& {Voyeux}}]{Pelouze2022}
{Pelouze}, G., {Auch{\`e}re}, F., {Bocchialini}, K., {et~al.} 2022, \aap, 658,
  A71

\bibitem[{{Petrova} {et~al.}(2023){Petrova}, {Magyar}, {Van Doorsselaere}, \&
  {Berghmans}}]{Petrova2023}
{Petrova}, E., {Magyar}, N., {Van Doorsselaere}, T., \& {Berghmans}, D. 2023,
  \apj, 946, 36

\bibitem[{{Reale}(2014)}]{Reale2014}
{Reale}, F. 2014, Living Reviews in Solar Physics, 11, 4

\bibitem[{{R{\'e}ville} {et~al.}(2020){R{\'e}ville}, {Velli}, {Panasenco},
  {Tenerani}, {Shi}, {Badman}, {Bale}, {Kasper}, {Stevens}, {Korreck},
  {Bonnell}, {Case}, {de Wit}, {Goetz}, {Harvey}, {Larson}, {Livi},
  {Malaspina}, {MacDowall}, {Pulupa}, \& {Whittlesey}}]{Reville2020}
{R{\'e}ville}, V., {Velli}, M., {Panasenco}, O., {et~al.} 2020, \apjs, 246, 24

\bibitem[{{Rosner} {et~al.}(1978){Rosner}, {Tucker}, \& {Vaiana}}]{Rosner1978}
{Rosner}, R., {Tucker}, W.~H., \& {Vaiana}, G.~S. 1978, ApJ, 220, 643

\bibitem[{{Serio} {et~al.}(1981){Serio}, {Peres}, {Vaiana}, {Golub}, \&
  {Rosner}}]{Serio1981}
{Serio}, S., {Peres}, G., {Vaiana}, G.~S., {Golub}, L., \& {Rosner}, R. 1981,
  ApJ, 243, 288

\bibitem[{{Shi} {et~al.}(2021){Shi}, {Van Doorsselaere}, {Guo}, {Karampelas},
  {Li}, \& {Antolin}}]{Shi2021}
{Shi}, M., {Van Doorsselaere}, T., {Guo}, M., {et~al.} 2021, \apj, 908, 233

\bibitem[{{Shi} {et~al.}(2024){Shi}, {Manchester}, {Landi}, {van der Holst},
  {Szente}, {Chen}, {T{\'o}th}, {Bertello}, \& {Pevtsov}}]{Shi2024}
{Shi}, T., {Manchester}, W., {Landi}, E., {et~al.} 2024, \apj, 961, 60

\bibitem[{{Shoda} {et~al.}(2019){Shoda}, {Suzuki}, {Asgari-Targhi}, \&
  {Yokoyama}}]{Shoda2019}
{Shoda}, M., {Suzuki}, T.~K., {Asgari-Targhi}, M., \& {Yokoyama}, T. 2019,
  \apjl, 880, L2

\bibitem[{{Shrivastav} {et~al.}(2024){Shrivastav}, {Pant}, {Berghmans},
  {Zhukov}, {Van Doorsselaere}, {Petrova}, {Banerjee}, {Lim}, \&
  {Verbeeck}}]{Shrivastav2024}
{Shrivastav}, A.~K., {Pant}, V., {Berghmans}, D., {et~al.} 2024, \aap, 685, A36

\bibitem[{{Shrivastav} {et~al.}(2025){Shrivastav}, {Pant}, {Kumar},
  {Berghmans}, {Van Doorsselaere}, {Banerjee}, {Petrova}, \&
  {Lim}}]{Shrivastav2025}
{Shrivastav}, A.~K., {Pant}, V., {Kumar}, R., {et~al.} 2025, \apj, 979, 6

\bibitem[{Tanaka(1994)}]{Tanaka1994}
Tanaka, T. 1994, Journal of Computational Physics, 111, 381

\bibitem[{{Terradas} {et~al.}(2008){Terradas}, {Andries}, {Goossens},
  {Arregui}, {Oliver}, \& {Ballester}}]{Terradas2008}
{Terradas}, J., {Andries}, J., {Goossens}, M., {et~al.} 2008, \apjl, 687, L115

\bibitem[{{Tiwari} {et~al.}(2021){Tiwari}, {Morton}, \&
  {McLaughlin}}]{Tiwari2021}
{Tiwari}, A.~K., {Morton}, R.~J., \& {McLaughlin}, J.~A. 2021, \apj, 919, 74

\bibitem[{{T{\'o}th} \& {Odstr{\v{c}}il}(1996)}]{Toth1996}
{T{\'o}th}, G. \& {Odstr{\v{c}}il}, D. 1996, Journal of Computational Physics,
  128, 82

\bibitem[{{van Ballegooijen} {et~al.}(2017){van Ballegooijen}, {Asgari-Targhi},
  \& {Voss}}]{vanBallegooijen2017}
{van Ballegooijen}, A.~A., {Asgari-Targhi}, M., \& {Voss}, A. 2017, \apj, 849,
  46

\bibitem[{{van der Holst} {et~al.}(2014){van der Holst}, {Sokolov}, {Meng},
  {Jin}, {Manchester}, {T{\'o}th}, \& {Gombosi}}]{vanderHolst2014}
{van der Holst}, B., {Sokolov}, I.~V., {Meng}, X., {et~al.} 2014, \apj, 782, 81

\bibitem[{{Van Doorsselaere} {et~al.}(2014){Van Doorsselaere}, {Gijsen},
  {Andries}, \& {Verth}}]{VanDoorsselaere2014}
{Van Doorsselaere}, T., {Gijsen}, S.~E., {Andries}, J., \& {Verth}, G. 2014,
  \apj, 795, 18

\bibitem[{{Van Doorsselaere} {et~al.}(2024){Van Doorsselaere}, {Magyar},
  {Sieyra}, \& {Goossens}}]{VanDoorsselaere2024}
{Van Doorsselaere}, T., {Magyar}, N., {Sieyra}, M.~V., \& {Goossens}, M. 2024,
  Journal of Plasma Physics, 90, 905900113

\bibitem[{{Van Doorsselaere} {et~al.}(2025){Van Doorsselaere}, {Sieyra},
  {Magyar}, {Goossens}, \& {Banovi{\'c}}}]{VanDoorsselaere2025}
{Van Doorsselaere}, T., {Sieyra}, M.~V., {Magyar}, N., {Goossens}, M., \&
  {Banovi{\'c}}, L. 2025, \aap, 696, A166

\bibitem[{{Van Doorsselaere} {et~al.}(2020){Van Doorsselaere}, {Srivastava},
  {Antolin}, {Magyar}, {Vasheghani Farahani}, {Tian}, {Kolotkov}, {Ofman},
  {Guo}, {Arregui}, {De Moortel}, \& {Pascoe}}]{VanDoorsselaere2020}
{Van Doorsselaere}, T., {Srivastava}, A.~K., {Antolin}, P., {et~al.} 2020,
  \ssr, 216, 140

\bibitem[{{Verdini} \& {Velli}(2007)}]{Verdini2007}
{Verdini}, A. \& {Velli}, M. 2007, \apj, 662, 669

\bibitem[{{Vesecky} {et~al.}(1979){Vesecky}, {Antiochos}, \&
  {Underwood}}]{Vesecky1979}
{Vesecky}, J.~F., {Antiochos}, S.~K., \& {Underwood}, J.~H. 1979, \apj, 233,
  987

\bibitem[{{Vidotto} {et~al.}(2023){Vidotto}, {Bourrier}, {Fares}, {Bellotti},
  {Donati}, {Petit}, {Hussain}, \& {Morin}}]{Vidotto2023}
{Vidotto}, A.~A., {Bourrier}, V., {Fares}, R., {et~al.} 2023, \aap, 678, A152

\bibitem[{{Washinoue} \& {Suzuki}(2023)}]{Washinoue2023}
{Washinoue}, H. \& {Suzuki}, T.~K. 2023, ApJ, 953, 74

\bibitem[{{Xia} {et~al.}(2018){Xia}, {Teunissen}, {El Mellah}, {Chan{\'e}}, \&
  {Keppens}}]{Xia2018}
{Xia}, C., {Teunissen}, J., {El Mellah}, I., {Chan{\'e}}, E., \& {Keppens}, R.
  2018, \apjs, 234, 30

\bibitem[{{Zhong} {et~al.}(2026){Zhong}, {Hillier}, \& {Arregui}}]{Zhong2026}
{Zhong}, S., {Hillier}, A., \& {Arregui}, I. 2026, \apj, 1000, 68

\end{thebibliography}

\begin{appendix}

\nolinenumbers

\section{Behaviour of the equations in the RTVS-based approach near the bottom boundary} \label{sect:app_numeric_bottom}

In this part, further reasoning is provided as to why the RTVS-based approach predicts the scaling law of $\bar{W}_\textrm{k} \propto T^{-1/2}$ to hold very well near the loop footpoints. Hence, the behaviour of the set of equations~\eqref{eq:numeric_Wk}, \eqref{eq:numeric_P} and \eqref{eq:numeric_T} in the limit of the bottom boundary (when $s\rightarrow 0$) is examined more closely. Only the scaling relationships between pressure, temperature and kink wave energy density are examined; we assume that all other parameters, namely the magnetic field, density contrast, filling factor and radius, can be taken to be constant in this limit.

We recall the boundary conditions at $s=0$, namely $P(0) = P_0$, $T(0)= T_0$ and $T'(0) = 0$. The latter two of these conditions mean that we can also treat the temperature as a constant, which, together with the fact that $\cos(0)=1$, implies that near the bottom boundary Eq.~\eqref{eq:numeric_P} reduces to
\begin{equation}
    P' = -\frac{\mu m_\textrm{p} g_0}{\nu k_\textrm{B}} \frac{P}{T_0},
\end{equation}
which means that
\begin{equation}
    (\ln P)' = -\frac{\mu m_\textrm{p} g_0}{\nu k_\textrm{B}T_0}
\end{equation}
and after integrating
\begin{equation}
    P = P_0 \exp\left(-\frac{\mu m_\textrm{p} g_0}{\nu k_\textrm{B}T_0}s\right).
\end{equation}

Next, we use this in Eq.~\eqref{eq:numeric_Wk_transformed} (which is equivalent to Eq.~\eqref{eq:numeric_Wk}) to get
\begin{equation}
    \left(\Tilde{S}^\pm\right)' \propto \left(\frac{P}{T_0}\right)^{1/4} \propto \frac{P_0}{T_0}\,\exp\left(-\frac{1}{4}\frac{\mu m_\textrm{p} g_0}{\nu k_\textrm{B}T_0}s\right),
\end{equation}
which after integrating leads to
\begin{equation}
    \Tilde{S}^\pm \propto \frac{-4\nu k_\textrm{B}}{\mu m_\textrm{p} g_0}\;P_0\exp\left(-\frac{1}{4}\frac{\mu m_\textrm{p} g_0}{\nu k_\textrm{B}T_0}s\right) \propto P^{1/4}.
\end{equation}
Substituting back to $W_\textrm{k}^\pm$ finally gives
\begin{equation}
    W_\textrm{k}^\pm \propto P^{-1/2} \left(Q^\mp\right)^{-1} \propto T_0^{-1/2},
\end{equation}
since $Q \propto \sqrt{T_0/P}$ from Eqs.~\eqref{eq:Q_1D} and \eqref{eq:alpha_kink}. This also means that $\bar{W}_\textrm{k} \propto T_0^{-1/2},$ which is the relationship visible in Fig.~\ref{fig:RTV_fitting} in the low-temperature regime. Even though this relationship is derived here under the assumption of constant temperature, this relationship seems to be satisfied by the RTVS-based model very well for the range of $T$ of more than one order of magnitude.

\section{Analytical fits to the kink wave energy density vs. temperature relationship obtained using the RTV-based approach}\label{sect:app_Wk_T_fitting}

\begin{figure}[ht]
    \centering
    \includegraphics[width=\hsize]{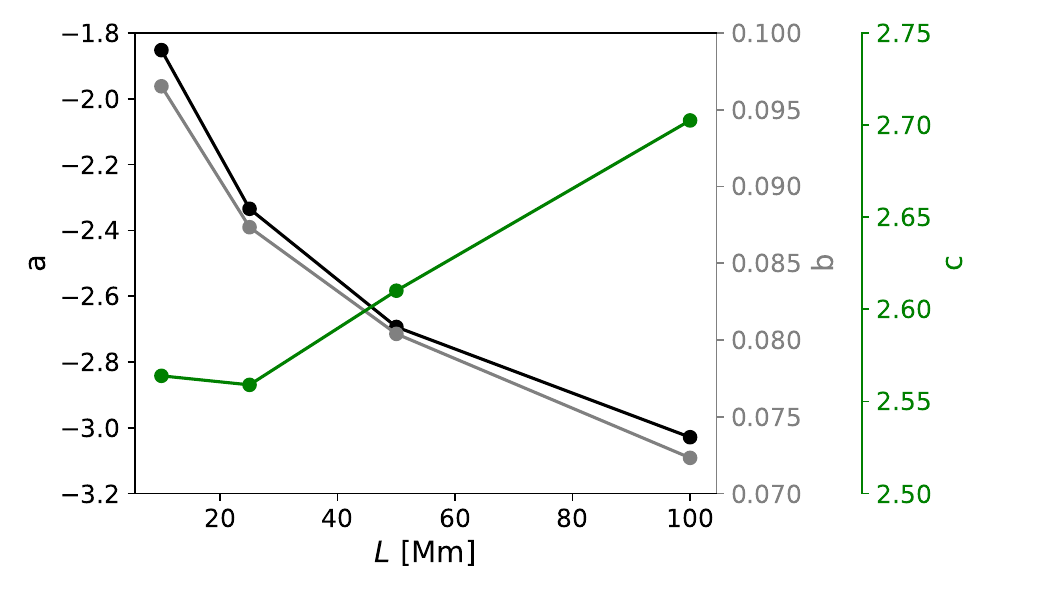}
    \caption{Fitted coefficients $a, b$ and $c$ for the loops with $L=10, 25, 50$ and $100$ Mm and $P_0=0.5$ dyn cm$^{-2}$. Different coefficients and corresponding vertical axes are shown in different colours.}
    \label{fig:RTV_fitting_coeff}
\end{figure}

\begin{figure}[ht]
    \centering
    \includegraphics[width=0.8\hsize]{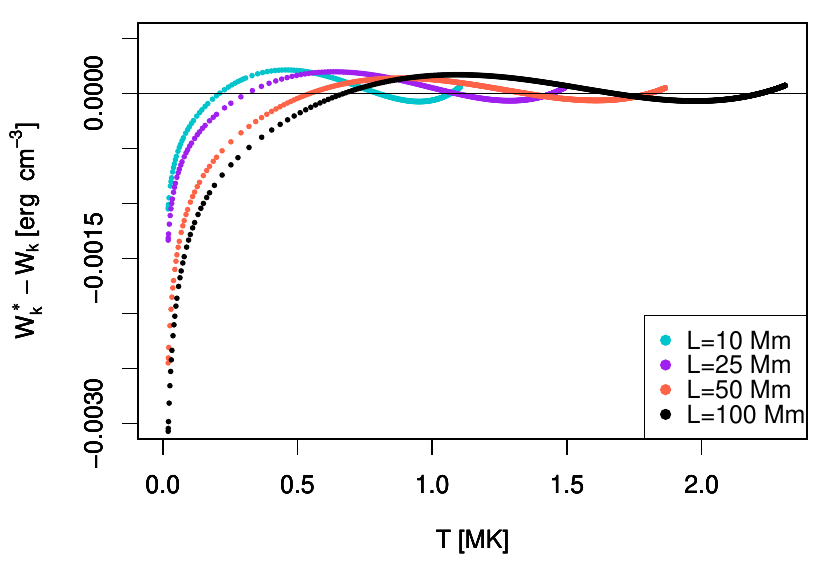}
    \caption{Error of the fit represented by Eq.~\eqref{eq:Wk_fitting} (denoted by $W_\textrm{k}^*$) to the results of the RTVS-based approach plotted against $T$ for different loop lengths.}
    \label{fig:RTV_fitting_residuals}
\end{figure}

As already mentioned in Sect.~\ref{sect:res_locality}, the relationship between $\bar{W}_\textrm{k}$ and $T$ obtained using the RTVS-based model can be approximated analytically. Specifically, we find that very good fits can be obtained using the following approximation, with $a, b$ and $c$ being free parameters:
\begin{equation}\label{eq:RTV_fitting_perfect_model}
    \ln \bar{W}_\textrm{k} = a - 0.5\ln T - bT^c,
\end{equation}
which corresponds to the following relationship between $\bar{W}_\textrm{k}$ and $T$: 
\begin{equation}
    \bar{W}_\textrm{k} = T^{-0.5} \exp(a-bT^c). \label{eq:Wk_fitting}
\end{equation}
The role of the additional exponential factor compared to the locality assumption is to follow the results obtained by the RTVS-based method more closely for higher temperatures, while at the same time leaving the dominant scaling relationship for the lower temperatures untouched. Parameters $a, b$ and $c$ for the same 4 loops shown in Fig.~\ref{fig:RTV_fitting} are given in Fig.~\ref{fig:RTV_fitting_coeff}. These parameters are obtained by means of non-linear least-squares fitting, which uses the Gauss-Newton algorithm. We also show the residuals of the fits in Fig.~\ref{fig:RTV_fitting_residuals}, confirming that the fits deviate from the numerical results by at most 1\% for all cases and throughout the loop (and most of the time less than that).

\section{Additional plots of scaling fits}\label{sect:app_scaling}

\begin{figure}[ht]
    \centering
    \subfloat[RTVS]{\includegraphics[width=0.8\hsize]{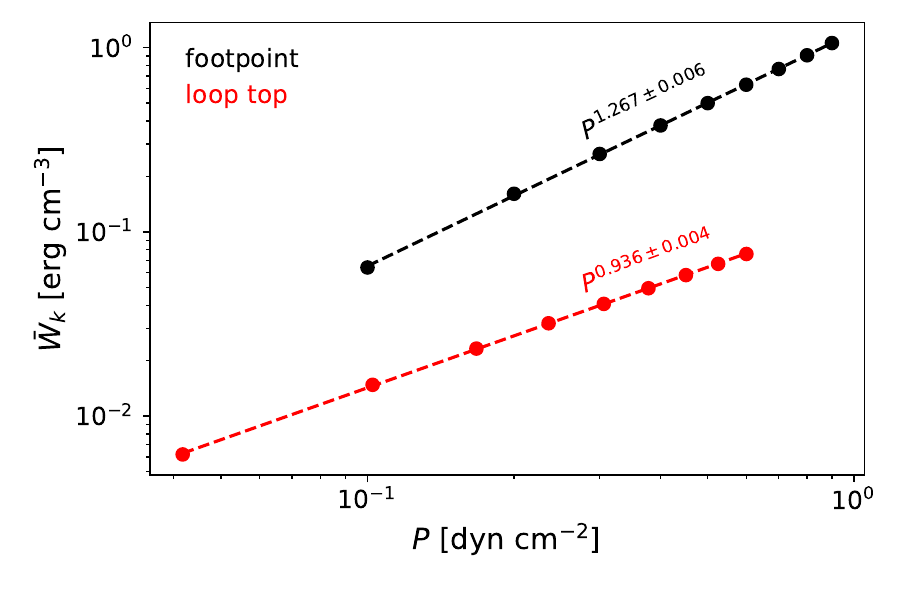}}
    
    \subfloat[UAWSoM]{\includegraphics[width=0.8\hsize]{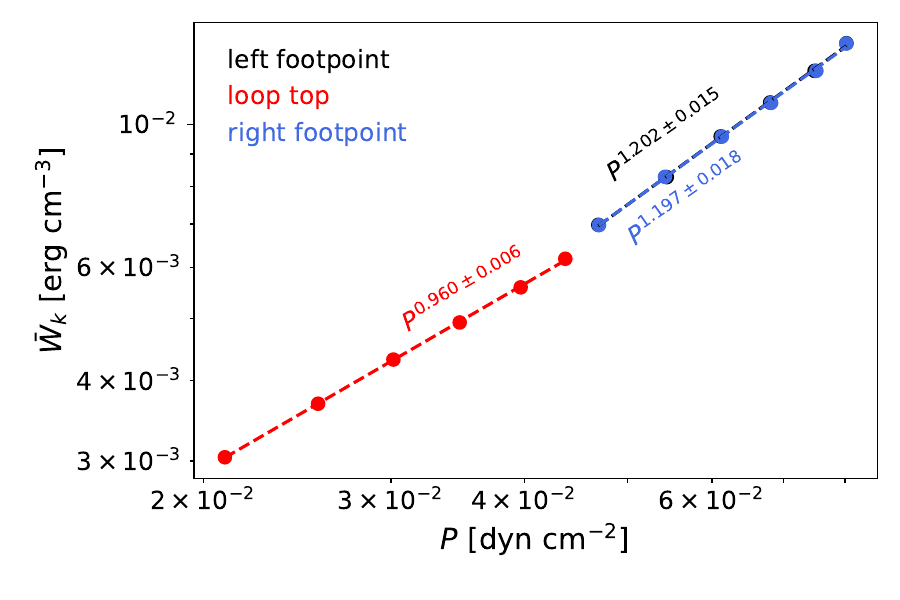}}

    \caption{Plots showing the scaling laws between $\bar{W}_\textrm{k}$ and $P$ for results obtained using the RTVS-based method (top) and \mbox{UAWSoM} (bottom). Fitted scaling laws are represented by dashed lines, each corresponding to the specific part of the loop. (Same as Fig.~\ref{fig:scaling_Wk_P}, only with $W_\textrm{k}^-$ being replaced by $\bar{W}_\textrm{k}$.)}
    \label{fig:scaling_app_Wktot_P} 
\end{figure}

\begin{figure}[ht]
    \centering
    \includegraphics[width=0.8\hsize]{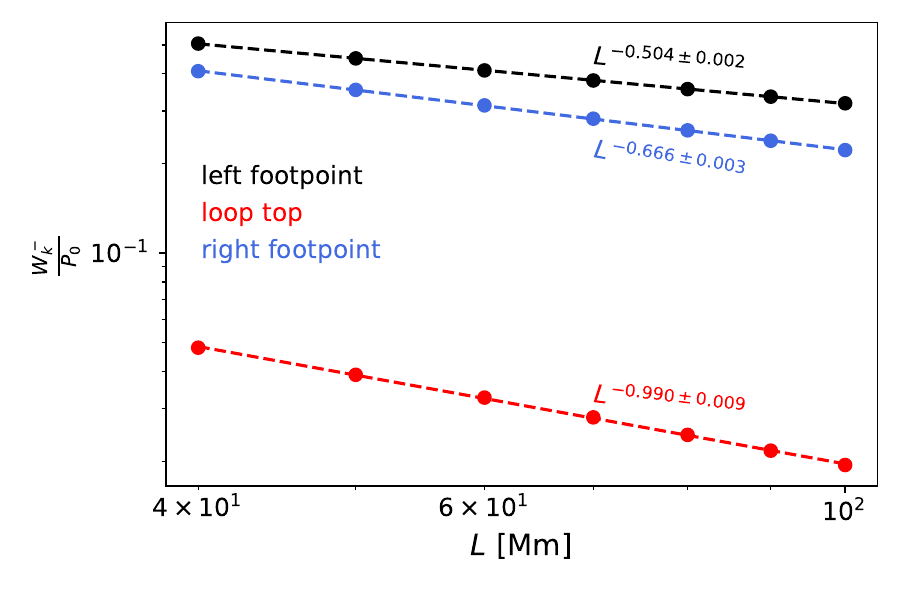}

    \caption{Data obtained by the RTVS-based model and the corresponding scaling fits between $W_\textrm{k}^-/P_0$ and $L$. Fits are represented by dashed lines, each corresponding to the specific part of the loop.}
    \label{fig:scaling_app1} 
\end{figure}

\begin{figure}[ht]
    \centering
    \subfloat[RTVS]{\includegraphics[width=0.8\hsize]{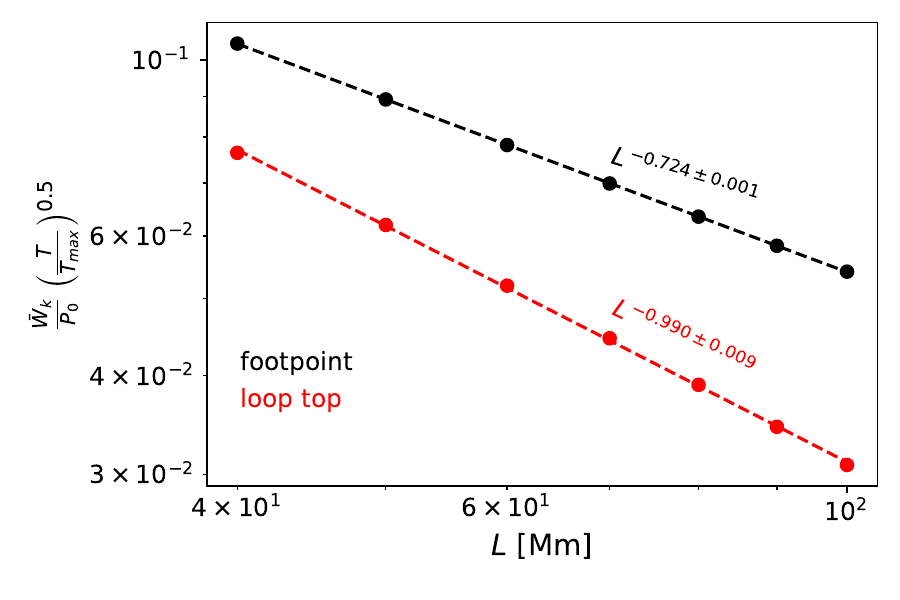}}
    
    \subfloat[UAWSoM]{\includegraphics[width=0.8\hsize]{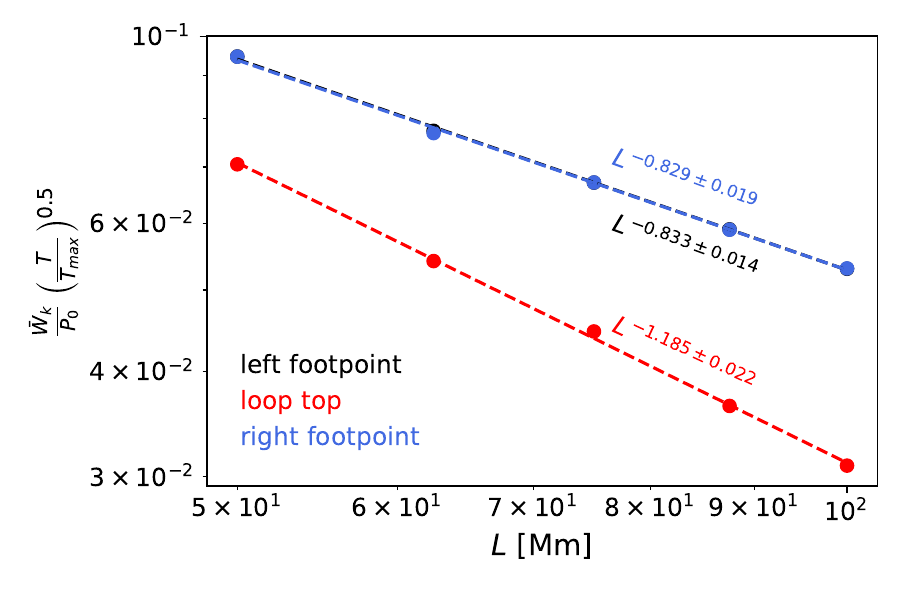}}

    \caption{Plots showing the scaling laws between appropriately scaled $\bar{W}_\textrm{k}$ (which was defined in Eq.~\eqref{eq:Wk_tot_definition}) and $L$ for results obtained using RTVS-based method (top) and \mbox{UAWSoM} (bottom). Fitted scaling laws are represented by dashed lines, each corresponding to the specific part of the loop.}
    \label{fig:scaling_app_Wktot_L} 
\end{figure}

In this section, we add the figures showing the scaling fits behind the scaling relationships which were only listed in the body of Sect.~\ref{sect:res_scalings} for completeness. The method used to estimate the scaling exponents in the figures here is the same as for the figures shown in Sect.~\ref{sect:res_scalings}. Figures~\ref{fig:scaling_app_Wktot_P}, \ref{fig:scaling_app1} and \ref{fig:scaling_app_Wktot_L} correspond to Eqs.~\eqref{eq:scaling_add1}, ~\eqref{eq:scaling_add2} and ~\eqref{eq:scaling_add3}, respectively.

\end{appendix}

\end{document}